\documentclass{aa}  

\usepackage{graphicx}
\usepackage{amsmath, bm}
\usepackage{ulem}  

\usepackage{txfonts}
\usepackage{xfrac}  

\usepackage{xcolor}  
\definecolor{aaBlue}{RGB}{0,0,200}  

\usepackage[colorlinks=true, 
            linkcolor=aaBlue, 
            citecolor=aaBlue, 
            urlcolor=aaBlue]{hyperref}

\begin{document}

   \title{Deep and Sparse Denoising Benchmarks for Spectral Data Cubes of High-z Galaxies: From Simulations to ALMA observations}

   \author{Arnab Lahiry
          \inst{1,2} \fnmsep\thanks{E-mail: arnablahiry08@gmail.com}
          , Tanio D\'iaz-Santos \inst{1}
          , Jean-Luc Starck\inst{1,3}
          , Niranjan Chandra Roy\inst{4}
          , Daniel Anglés-Alcázar\inst{4}
          , Grigorios Tsagkatakis\inst{1,2}
          , and Panagiotis Tsakalides\inst{1,2}
          }

       \institute{Institutes of Computer Science and Astrophysics, Foundation for Research and Technology Hellas (FORTH), 100 Nikolaou Plastira str., Vassilika Vouton, Heraklion, 70013, Greece
         \and
            Departments of Physics and Computer Science, University of Crete, Voutes Campus, Vasilika Voutes, Heraklion, 70013, Greece
         \and
             Université Paris-Saclay, Université Paris Cité, CEA, CNRS, AIM, 91191, Gif-sur-Yvette, France
         \and
             Department of Physics, University of Connecticut, 196 Auditorium Road, U-3046, Storrs, CT 06269, USA}

   \date{Received September 15, 1996; accepted March 16, 1997}


\abstract
{Beyond cosmic noon, galaxies usually appear as faint whispers amid overwhelming noise, yet this epoch is key to understanding
massive galaxy assembly. ALMA’s sensitivity to cold dust and [\ion{C}{II}] emission allows us to probe their interstellar medium, but faint signals are still challenging, rendering robust denoising essential.}
{We evaluate denoising strategies, including classical statistical methods, sparse unsupervised representations, and supervised deep learning, to identify techniques that suppress noise while preserving flux and spectral-spatial morphology.}
{We develop a physically motivated synthetic dataset of spectral cubes simulating rotating disk galaxies for training and evaluation. We benchmark Principal Component Analysis (PCA), Independent Component Analysis (ICA), iterative soft thresholding with 2D-1D wavelets (IST), and a supervised 3D U-Net across peak SNRs of $\sim$2.5--8, applied to (i) toy cubes, (ii) synthetic [\ion{C}{II}] IFU cubes from FIRE simulations, and (iii) ALMA observations of $z\sim5$ galaxies from the CRISTAL sample and the quasar W2246$-$0526. Performance is assessed via RMSE, flux conservation, morphology, and SNR improvement.}
{PCA and ICA provide limited noise reduction and struggle with correlated noise. IST reduces noise at moderate SNRs but can suppress emission at low SNRs. The 3D U-Net outperforms IST on synthetic cubes, particularly at low SNR, though it may overestimate flux or hallucinate faint structures in this regime. On high SNR real data with relatively simple morphologies, the U-Net and IST achieve comparable performance. However, on low SNR real data with complex morphologies not represented in the training set, the U-Net underperforms relative to IST, highlighting the challenges of generalization beyond the training distribution. In ALMA-CRISTAL cubes, both IST and U-Net conserve $>91\%$ of flux and increase SNR by $>6$. For the extreme case of W2246--0526, the U-Net recovers $\sim80\%$ of flux at moderate SNR, whereas IST robustly conserves flux and improves SNR by $\sim3$.}
{Deep learning trained on synthetic data generalizes effectively, though flux bias and interpretability challenges remain at low SNR. The addition of physically motivated priors and uncertainty quantification will enhance robustness. This framework of synthetic, simulated, and real datasets offers a pathway for transferable denoising in surveys with ALMA, VLT/MUSE, and JWST IFUs.}

   \keywords{infrared: galaxies--
            galaxies: high-redshift --
            techniques: image spectroscopy --
            methods: data analysis --
            methods: statistical}

\authorrunning{Lahiry et al.}
\titlerunning{Deep and Sparse Denoising Benchmarks of Spectral Data Cubes of High-z Galaxies}
\maketitle


\section{Introduction}

Denoising is a critical step in the analysis of astrophysical spectral cubes, particularly for high-redshift ($z>3$) sources where galaxies appear faint and the data are dominated by noise. This epoch, beyond the peak of cosmic star formation (“cosmic noon”), marks a pivotal phase in the assembly and quenching of massive galaxies. Extracting faint, spatially and spectrally resolved signals at these redshifts is essential to uncovering the physical processes that shape early galaxy evolution, including star formation, gas accretion, and feedback \citep{Carilli2013}.  

The Atacama Large Millimeter/submillimeter Array (ALMA) is uniquely suited to probe the cold gas and dust in the early Universe. Its sensitivity to the [C\,\textsc{ii}] 158\,$\mu$m fine-structure line and dust continuum provides an unparalleled view of the interstellar medium (ISM) and obscured star formation at $z>3$ \citep{Carilli2013,Lagache2018}. Indeed, ALMA has revealed [C\,\textsc{ii}] and dust emission from galaxies out to $z\sim8$, enabling studies of their morphology, kinematics, and ISM properties at early cosmic times \citep{Bethermin2020, Tsukui2021, herrera}.  

Yet, observations at $z>3$ remain challenged by low signal-to-noise ratios (SNRs) and correlated noise introduced by the interferometric beam. For example, the ALPINE survey detected [C\,\textsc{ii}] in only $\sim64\%$ of its sample of $4<z<6$ galaxies at a conservative $3.5\sigma$ threshold, with a median SNR of $\sim6$ and many sources falling below detection limits \citep{Bethermin2020}. Spatially correlated noise further complicates the reliable extraction of signals \citep{Tsukui2023}. Robust denoising methods capable of suppressing both uncorrelated and beam-correlated noise are therefore essential.

Recovering both resolved and unresolved emission is crucial for addressing key scientific questions. Spatially resolved [C\,\textsc{ii}] maps reveal the geometry and extent of star formation, with observations uncovering disks, clumps, and even spiral arms in $z>4$ galaxies \citep{lee}. Resolved velocity fields provide constraints on dynamical masses, rotation curves, and feedback processes, including large-scale outflows \citep{Carilli2013,Carniani2020}. At the same time, many high-redshift galaxies remain unresolved or barely resolved at ALMA’s typical resolution ($\sim0.5''$), and their integrated [C\,\textsc{ii}] and dust fluxes, line widths, and velocity offsets still yield key insights into their star formation rates, ISM conditions, and kinematics \citep{scoville, telikova, birkin,li},. Finally, the ratio of [C\,\textsc{ii}] and IR luminosities, or [O\,\textsc{iii}]/[C\,\textsc{ii}] ratios, informs models of metallicity, ISM density, and radiation field strength \citep{Ferrara2019,Carniani2020}.

To tackle these challenges, various denoising strategies have been developed and used for astrophysical data. Classical statistical methods such as Principal Component Analysis (PCA) \citep{Hotelling_PCA, Jolliffe_PCA, Pearson_PCA} and Independent Component Analysis (ICA) \citep{Comon1994, Hyvarinen2001} rely on linear decompositions to separate signal from noise. Wavelet-based methods exploit multiscale, sparse representations to preserve features across spatial and spectral dimensions. More recently, supervised deep learning approaches, such as convolutional neural networks with U-Net architectures \citep{ronneberger2015u}, have shown promise by learning complex, nonlinear mappings directly from data \citep{Guglielmetti2024}.  

In this work, we present a comprehensive evaluation of these denoising strategies, applied to spectral cubes spanning increasing realism: (i) ALMA-like cubes generated from synthetic toy models of rotating disk galaxies, (ii) synthetic integral field unit (IFU) cubes from FIRE cosmological simulations, and (iii) [C\,\textsc{ii}] ALMA observations from the CRISTAL survey and the hyperluminous quasar W2246$-$0526. Our synthetic dataset is physically motivated and highly configurable, enabling controlled experiments across a range of noise levels, spatial resolutions, and source morphologies.  

By benchmarking PCA, ICA, iterative 2D–1D wavelet soft thresholding, and a supervised 3D U-Net, we systematically assess their ability to suppress noise while preserving astrophysical signals. Our results highlight the strengths and limitations of each method and demonstrate the promise of synthetic training sets for improving supervised denoising of complex, high-redshift observations.  

Since this work compares using a neural net with an encoding structure to PCA/ICA methods, it is worth noting the foundational comparison between autoencoders and principal component analysis established by \citet{hinton2006}.

While \citet{Guglielmetti2024} and similar works have generally focused on single-channel imaging, their strong results with U-Net-based models strengthen the case for the methodology developed in this work and also point the way to future improvements. For instance, recent advancements have addressed general correlated noise using U-Net-based diffusion models \citep{legin2023}. Similarly, joint-correlated-noise and PSF deconvolution has been explored by \citet{adam2025}. In the specific context of interferometry, methods for imaging in the visibility space using U-Net diffusion models have been proposed by \citet{feng2024} and \citet{dia2025}. Furthermore, for handling correlated noise in interferometry data converted to dirty images, recent works by \citet{connor2022}, \citet{terris2025}, and \citet{mars2025} offer relevant benchmarks.

As ALMA, JWST, and future surveys push deeper into the early Universe, robust denoising techniques will be indispensable for uncovering the faint signals that illuminate the physics of galaxy assembly and feedback. The adaptation and future development of aforementioned advanced architectures suggests that addressing correlated noise from single-channel images to 3D spectral cubes is a timely and necessary progression. The framework developed here provides a pathway toward reliable, generalizable denoising pipelines tailored for modern astrophysical data.

In the following sections, we detail the full experimental pipeline and analyses undertaken.  
\textbf{Section~2} describes the datasets used: our physically motivated synthetic toy data cubes, realistic mock IFU cubes from cosmological FIRE simulations, and ALMA observations of W2246–0526.  
\textbf{Section~3} introduces the evaluation criteria and performance metrics, including how we identify emission regions and quantify denoising performance.  
\textbf{Section~4} presents the unsupervised denoising methods - PCA, ICA, and iterative 2D-1D wavelet thresholding, along with their theoretical foundations and implementation details.  
\textbf{Section~5} describes the supervised deep-learning approach using a 3D U-Net, including the network architecture, training procedure, and training data.  
\textbf{Section~6} reports the results of benchmarking these methods on synthetic, simulated, and real datasets, analyzing their performance across varying SNRs and morphologies.  
\textbf{Section~7} concludes with a discussion of the implications of our findings, current limitations, and directions for future work.

\begin{table*}[h!]
\centering
\begin{tabular}{|l|l|l|l|}
\hline
\textbf{Parameter Description}                         & \textbf{Notation}       & \textbf{Range}                                                                                                                                            & \textbf{Units} \\ \hline
Number of companion galaxies          & $N_{\rm gals}$              & $\sim \mathcal{U}_d(\{1, 2, 3, 4\})$                                                                                                                   & --              \\ \hline
Distance of companion galaxies from the central galaxy & $d_{\rm gal}$               & $\sim \mathcal{U}(0, 20)$                                                                                                                                 & kpc            \\ \hline
Effective/half-light radius (in the plane of the disk) & $R_{\rm e}$                   & \begin{tabular}[c]{@{}l@{}}Central galaxy: $\sim \mathcal{U}(4, 6)$\\ Companion galaxies: $\sim \mathcal{U}(1, 3)$\end{tabular}      & kpc            \\ \hline
Scale height & $h_z$                   & \begin{tabular}[c]{@{}l@{}}Central galaxy: $\sim \mathcal{U}(0.5, 1)$\\ Companion galaxies: $\sim \mathcal{U}(0.15,0.5)$\end{tabular}      & kpc            \\ \hline
S\'ersic index                          & $n$                     & $\sim \mathcal{U}(0.5, 1.5)$                                                                                                                              & --              \\ \hline
Effective/half-light flux density       & $S_{\rm e}$                   & \begin{tabular}[c]{@{}l@{}}Central galaxy: $\sim \mathcal{U}(0.08, 0.12)$\\ Companion galaxies: $\sim \mathcal{U}(0.02, 0.06)$\end{tabular} & $\mathrm{Jy\,px^{-1}}$         \\ \hline
Field of view (X and Y) - whole system        & $FOV$               & $\sim \mathcal{U}(46, 576)$           & kpc         \\ \hline

Angles of inclination around the X and Y axes          & $\theta_x, \theta_y$    & $\sim \mathcal{U}(-180, 180)$                                                                                                                                & degrees        \\ \hline
Initial velocity dispersion along the line of sight (Z axis)    & $\sigma_{v_z}$          & $\sim \mathcal{U}(10, 30)$                                                                                                                                &  km\,s$^{-1}$    \\ \hline
Peak signal-to-noise ratio for spatially correlated Gaussian noise                    & $\text{SNR}_{\rm peak}$ & $\sim \mathcal{U}(2.5, 8)$                                                                                                                               & --              \\ \hline
\end{tabular}

\vspace{0.8em}
\caption{Description and sampled ranges of the variable parameters used to construct a dataset of multiple spectral cubes.}
\label{table:params}
\end{table*}

\section{Data}

\subsection{Toy Data of Rotating Galaxies}\label{subsection:toy}

We construct synthetic spectral cubes designed to simulate observed \emph{flux density} of rotating disk galaxies, providing a controlled environment for testing denoising algorithms and kinematic extraction under realistic noise and beam conditions. The mock cubes consist of a bright central galaxy and several fainter satellites, each specified by structural and dynamical parameters. We aim to create a large, variable dataset designed for the evaluation of denoising methodologies, and primarily to serve as a robust dataset for training and validating our U-Net-based denoiser (See section \ref{sec:unet_denoising})

We first generate the galaxies’ three-dimensional light distribution directly in observed \textit{flux density per voxel} (Jy/voxel). We aim to simulate ALMA-like spectral cubes,
where we typically obtain the observed frequency of the emission line, and we can calculate the line-of-sight velocities for each spectral observation using the Doppler shift with respect to the central/systemic frequency of the emission line, scaled by the speed of light. However, for our toy cubes, we directly calculate the line-of-sight velocity by simulating realistic rotational velocities in the planes of the disks, rotating the whole system in three-dimensional space, and choosing a line-of-sight as the observed velocity axis. We then calculate projections of the flux density, binned along that line of sight.

The resulting cube is expressed as \emph{flux density per pixel} (Jy/pixel) in each spectral slice.  Finally, the cube is convolved with a Gaussian beam to transform it into units of \emph{flux density per beam} (Jy/beam), and beam-convolved spatially correlated Gaussian noise is added to simulate observational uncertainties.

We note that although high-redshift galaxies exhibit a vast range of complex morphologies, our synthetic generation pipeline adopts a simplified rotating disk model as the fundamental archetype. We acknowledge that this assumption introduces a morphological bias, as the denoising algorithms validated on this dataset are optimized for rotation-dominated kinematics. This stands as a limitation of the current physical model, as the exclusion of features such as extended diffuse emission, chaotic mergers, and gravitational lensing restricts immediate generalization to such targets. However, we posit that this dataset serves as an excellent foundational baseline for \textit{transfer learning}. A model pre-trained on these simplified simulations can capture fundamental spectral-spatial correlations, effectively serving as a "coarse" encoder that can subsequently be fine-tuned using smaller, specialized datasets incorporating detailed physical priors such as lensing effects or complex turbulence---thereby adapting the denoiser to specific observational tasks.

The construction pipeline comprises the following steps.

\subsection{Spatial Structure}
\label{sec:spatial_structure}
We define the intrinsic three-dimensional flux density per voxel as a product of a radial Sérsic profile (in the plane of the disk) and a vertical exponential profile:
\begin{equation}
S(x,y,z) =
S_e \cdot
\exp\left[-b_n \left( \left( \frac{\sqrt{x^2+y^2}}{R_e} \right)^{1/n} -1 \right)\right] \cdot
\exp\left(-\frac{|z|}{h_z}\right),
\end{equation}
where:
\begin{itemize}
\item $R_e$ is the effective (half-light) radius (in kpc),
\item $S_e$ is the flux density per voxel at $R_e$ (in Jy/voxel),
\item $n$ is the Sérsic index controlling the concentration of light,
\item $h_z$ is the vertical scale height (in kpc),
\item $b_n$ is a scaling constant ensuring that $R_e$ encloses half of the total light, computed by a polynomial approximation valid for $n > 0.36$: \citep{b_n_analytic}
\end{itemize}

\begin{equation}
b_n =
2n - \frac{1}{3} +
\frac{4}{405n} +
\frac{46}{25515n^2} +
\frac{131}{1148175n^3} -
\frac{2194697}{30690717750n^4}.
\end{equation}

The Sérsic profile determines the radial light distribution in the $X$-$Y$ plane, while the exponential profile governs the decay along the $Z$-axis, yielding a disk with finite thickness.  
This three-dimensional flux density field is defined on a Cartesian grid $(X,Y,Z)$, where the $X$-$Y$ plane corresponds to the galaxy disk and $Z$ is perpendicular to it, resulting in a static, face-on realisation of the galaxy.

\subsubsection{Kinematic Structure}

The rotational velocity of the disk is defined as a function of the galactocentric cylindrical radius $R = \sqrt{x^2 + y^2}$, assuming circular orbits in the plane of the disk. We adopt an empirical model based on the Milky Way’s rotation curve \citep{milky_way_rot_curve}:
 
\begin{equation}
v(R) = v_0 \cdot 1.022 \left(\frac{R}{R_0}\right)^{0.0803},
\end{equation}

\noindent where we vary the characteristic rotation velocity and radius
around standard values of $v_0 = 240 \text{ km/s}$ and $R_0 = 8.34 \text{ kpc}$.

The velocity vector at each point in the disk is oriented tangentially to the position vector ($\bm{r} = (x, y, 0)$) from the center of the galaxy to the voxel in the plane of the disk. The rotation direction is defined as the unit vector of the cross product:
\begin{equation}
\begin{aligned}
\bm{\hat{v}} &= \frac{\bm{r} \times \bm{\hat{z}}}{\|\bm{r}\|},& 
\bm{v} = v(R) \cdot \bm{\hat{v}}
\end{aligned}
\end{equation}

where $\bm{\hat{z}}$ is the unit vector along the vertical axis and $\|\bm{r}\| = R$.

To mimic realistic observations, we introduce a vertical velocity dispersion component $v_z$.

$\overline{v_z} = 0$, but we perturb it by adding Gaussian noise
\[
v_z \sim \mathcal{N}(0, \sigma_z), \quad \text{where } \sigma_z \approx 0.1 \cdot \max(v(R))
\]
This adds mild turbulence or pressure support typical in disk galaxies.

\begin{figure}[t!]
    \includegraphics[width=\columnwidth]{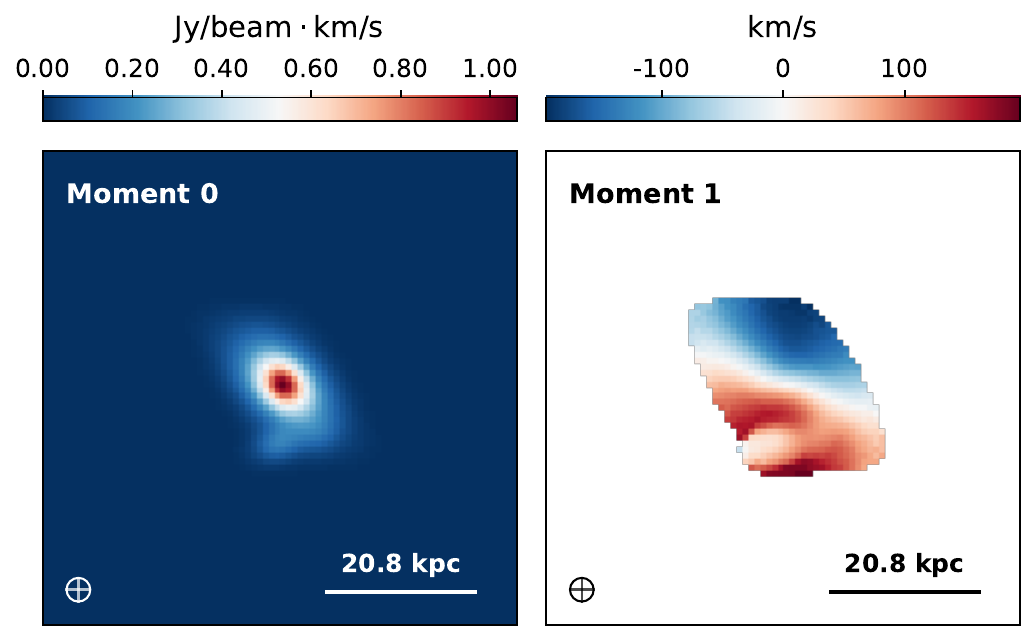}
    \includegraphics[width=\columnwidth]{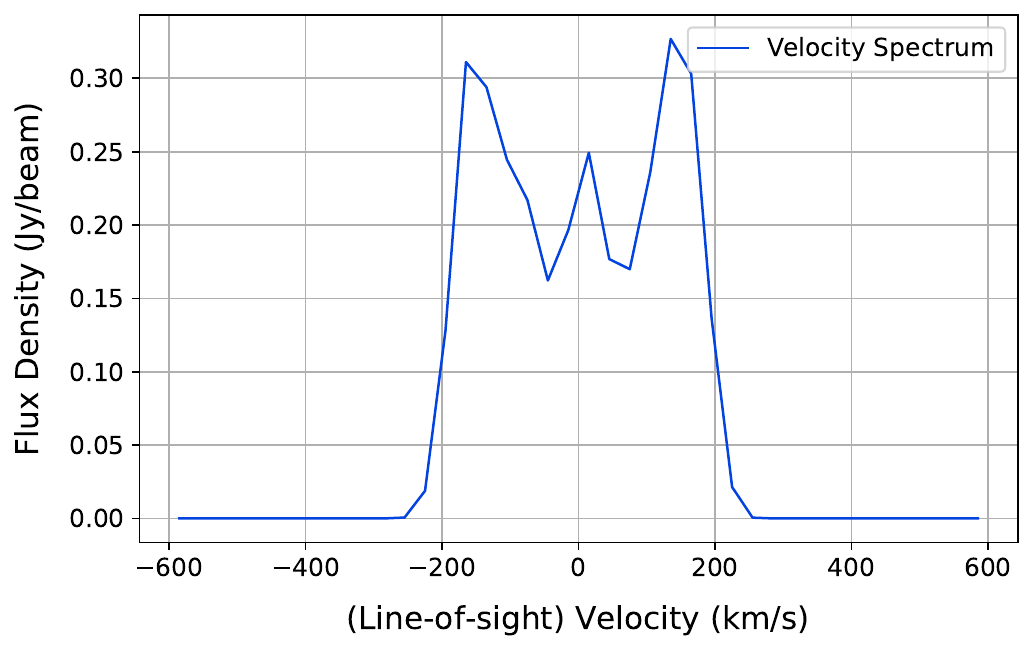}
    \caption{\textbf{Top Left:} Moment 0 map (integrated flux along line of sight) of a system with one large central galaxy and two satellites, both inclined uniquely in the field of view, and convolved with a two-dimensional Gaussian beam with FWHM = 3.75 px $\sim 3.12 \rm\: kpc$; \textbf{Top Right:} Moment 1 map (intensity weighted average velocity along the line of sight for each spaxel) within the emission region depicting the kinematics of the system; \textbf{Bottom:} Spatially integrated line-of-sight velocity spectrum of the system within the emission region, depicting a \textbf{double-horned spectral morphology} due to the distinctly modeled kinematic features due to the rotation of the galaxies.}
    \label{fig:moment_maps_spectrum_resolved}
\end{figure}

\begin{figure}[t!]
    \includegraphics[width=\columnwidth]{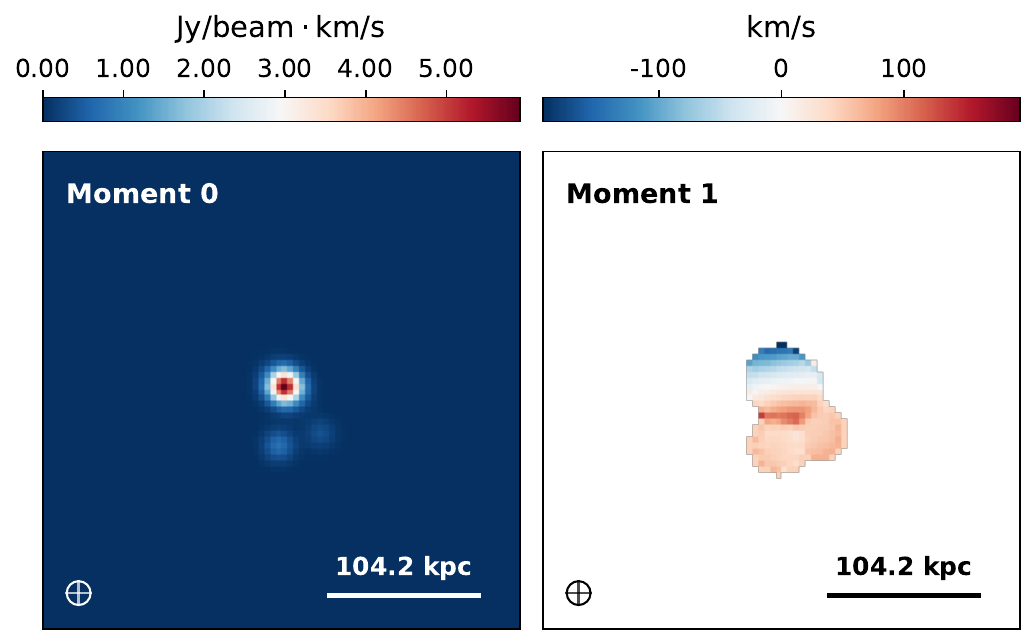}
    \includegraphics[width=\columnwidth]{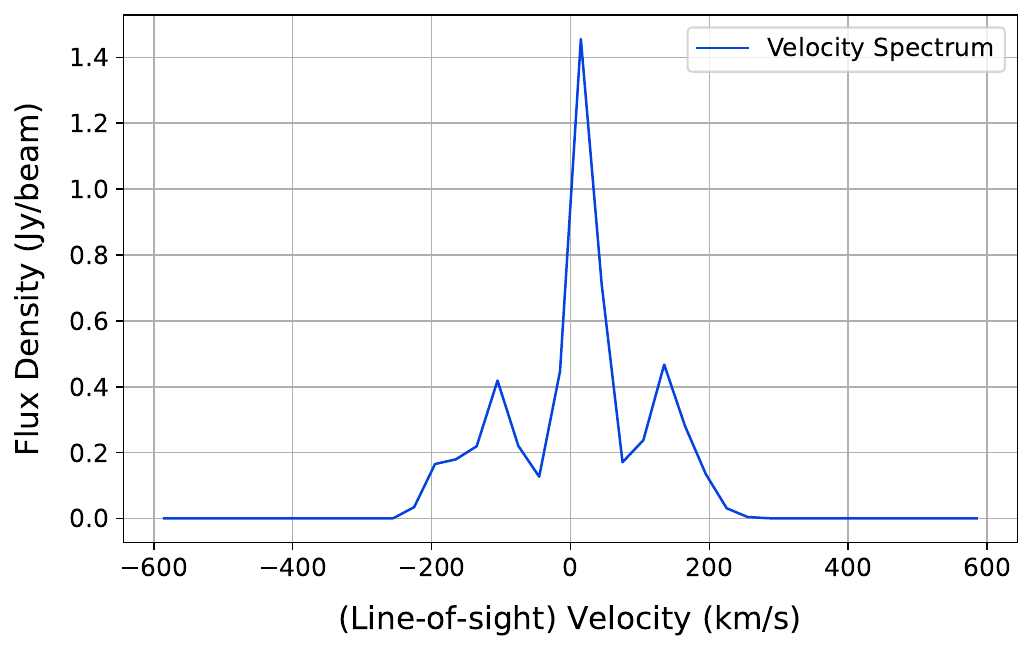}
    \caption{Simulation results for an \textbf{unresolved} system of galaxies in a larger field of view, convolved with a two-dimensional Gaussian beam with FWHM = 3.75px  $\sim 15.63 \rm \: kpc$. \textbf{Top Left:} Moment 0 map \textbf{Top Right:} Moment 1 map depicting the kinematics of the system; \textbf{Bottom:} Line-of-sight velocity spectrum of the system within the emission region, depicting a \textbf{peak at the central velocity} and non-substantially modeled kinematic features on either side due to the small size of the source in pixels within the field of view.}
    \label{fig:moment_maps_spectrum_unresolved}
\end{figure}

\subsubsection{System Rotation}

Galaxies are not always observed face-on in the sky. Their inclination significantly alters the projected appearance and observed line-of-sight velocity profiles. A face-on galaxy appears circular with little velocity gradient along the line of sight, while an edge-on system appears elongated with strong Doppler shifts.

To model arbitrary viewing angles of galaxies in the 3D universe, we rotate each galaxy’s 3D flux density and velocity fields using a composite rotation defined by two angles: $\theta_x$ and $\theta_y$.
    
We perform the following sequence of anti-clockwise rotations (1) around the \textbf{X-axis} by angle $\theta_x$: This tilts the galaxy away from the observer, simulating inclination; and (2) around the \textbf{Y-axis} by angle $\theta_y$: This introduces azimuthal variation in the viewing angle.

\begin{itemize}
    \item For the \textbf{flux density field}: We implement three-dimensional rotation of the complete spatial grid, preserving array dimensions and interpolating values between voxels to ensure consistency.
    
    \item For the \textbf{velocity field}: Each velocity vector $\bm{v}$ in the updated spatial grid is explicitly rotated:
    \[
    \bm{v}' = R \cdot \bm{v}.
    \]

    where, $R = R_Y(\theta_y) \cdot R_X(\theta_x)$, and $R_X$ and $R_Y$ are the three dimensional rotation matrices around the X and Y axes respectively.
\end{itemize}

This ensures the directional integrity of the motion is preserved in the inclined frame. After rotation, the $z$-component ($v_z$) of each velocity vector is chosen to be the line-of-sight velocity.

\subsubsection{Spectral Cube Construction}

For systems in which there is more than one galaxy, we have $n_{gal}$ number of three-dimensional spatial grids of individually rotated galaxies and calculated line-of-sight velocities. Each galaxy’s flux and velocity fields are embedded into a larger spatial grid representing the field of view of the full system. The central galaxy is situated at the center of the field of view, and the satellite galaxies are situated at randomly chosen positions within a controllable offset distance $d_{gal}$ from the central galaxy.

We define $n_s$ discrete velocity channels between the minimum and maximum line-of-sight velocities $v_{z,\min}$ and $v_{z,\max}$. These boundaries are set on the basis of the global velocity distribution across all galaxies. Each channel represents a small bin of velocities:
\[
\Delta v = \frac{v_{z,\rm obs,\max} - v_{z,\rm obs,\min}}{n_s}.
\]

For each velocity slice $v_i \in [v_{z,\rm obs,\min}, v_{z,\rm obs, \max}]$, we identify the voxels in the rotated spatial grid containing the flux density per voxel, that have a line of sight velocity $v_{z, \rm obs} \in [v_i, v_i + \Delta v]$, and we then integrate their flux densities along the line of sight, which in our case is the Z axis. This yields a two-dimensional image corresponding to the flux density per pixel, observed in the field of view at that particular line of sight velocity slice.

ALMA observations are convolved with the synthesized beam, which acts as the effective point spread function (PSF) and is expressed as a two-dimensional Gaussian. This induces spatial correlations over the beam area, making neighboring pixels statistically dependent. The beam sets the smallest resolvable scale, while the pixel grid samples the convolved signal; increasing pixel resolution does not add independent information, as pixels within a beam are fully correlated. 

To simulate these conditions, and for simplicity, we work with a general circular beam, 

To simulate these conditions, we employ a generalized circular Gaussian beam PSF for convolution to remain agnostic to the specific ellipticities and position angles that vary across different observations, thereby preventing unsupervised methods trained on the model from overfitting to a fixed beam geometry. We convolve the PSF with the projected two-dimensional flux density per velocity bin, and express it in units of Jy/beam by scaling with the area spanned by the beam in $\rm px^2$.

While this approximation is adequate for the benchmarks presented in this work, it nonetheless simplifies the intricate sidelobe structure characteristic of real interferometric PSFs. Future work that incorporates instrument or task-specific PSFs into dataset construction or fine-tuning could further improve the generalisation capabilities of models trained on this dataset.

The result is a spectral cube $X(x, y, v_z)$ that (1) encodes spatial morphology and line-of-sight kinematics, (2) includes multiple galaxies at arbitrary locations and inclinations, and (4) provides ground truth information for both flux density and spectra.

\subsubsection{A large variable dataset}\label{sec:variable_dataset}

\begin{figure*}[t!]
    \includegraphics[width=\textwidth]{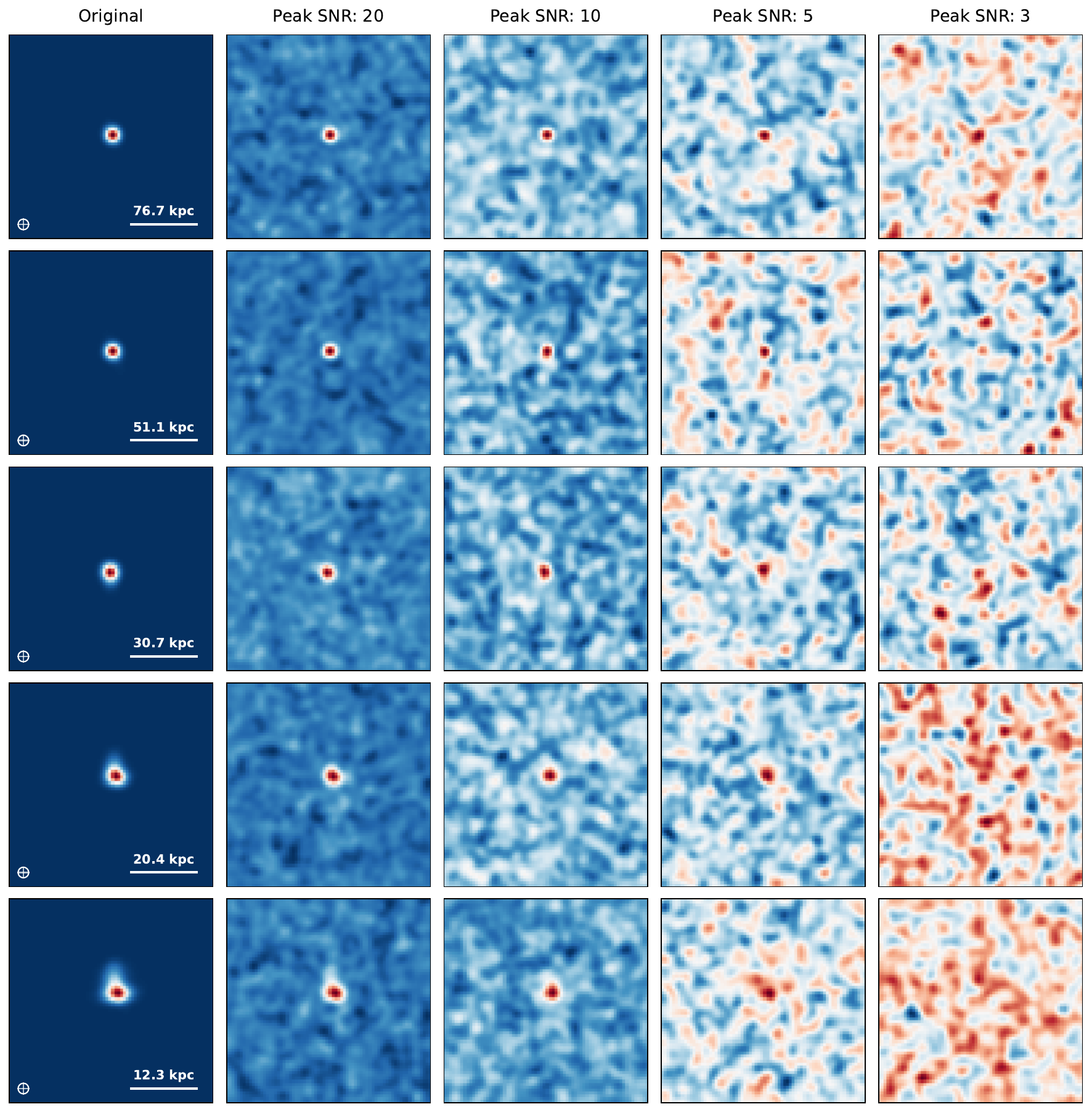}
    \caption{Visualisation of one spectral slice of a system with one galaxy at varying resolutions and levels of spatially correlated Gaussian noise.}
    \label{fig:toy_visualise}
\end{figure*}

The toy dataset is constructed from purely theoretical galaxy models, where every parameter is fully user-controlled. Varying these parameters generates unique spectral cubes, which is especially advantageous for large-scale comparative studies and supervised machine learning applications (Section \ref{sec:unet_denoising}). We sample the parameters defining the individual galaxies and the system from ranges detailed in table \ref{table:params}. 

We create a data set comprising $N_{\mathrm{cubes}}$ unique spectral cubes, each of dimensions $36 \text{ (spectral) } \times 72 \times 72 \text{ (spatial)}$ pixels. The data set includes systems with central galaxies spanning a range of effective radii, fields of view and inclinations.

For the analysis of denoising performance, we aim to compare results for sources in two broad classes of spectral cubes representing similarly intrinsically physically sized galaxies in kpc, but observed in different fields of view, with a fixed circular Gaussian beam of constant full width at half maximum (FWHM) in pixel units (3.75 px) - The central galaxy is considered to be \textbf{spatially resolved} by the beam if the effective diameter of the galaxy in pixel units is larger than the FWHM of the beam in the same units; the galaxy is \textbf{spatially unresolved} by the beam if the effective diameter is smaller than the FWHM of the beam in pixel units, resulting in compact emissions in a large field of view.

Figures \ref{fig:moment_maps_spectrum_resolved} and \ref{fig:moment_maps_spectrum_unresolved} show examples of a resolved and unresolved spectral cube of a system of galaxies and the corresponding moment maps - the moment 0 map: the integrated flux density map along the spectral axis, depicting the total emission per beam; and the moment 1 map: the flux density weighted velocity map depicting the weighted average line-of-sight velocity of the emission per beam, and the line-of-sight spatially integrated velocity spectra.

To generate inputs for denoising, synthetic noise is added to each spectral cube to mimic observational conditions with varying peak signal-to-noise ratios (SNRs), defined as the ratio of the maximum flux density to the noise standard deviation. For this study, we focus on a high-noise regime with peak SNRs uniformly sampled between 2.5 and 8 across all $N_{\mathrm{cubes}}$.

In ALMA observations, the noise, along with the signal, is convolved by the beam. To replicate these conditions in our simulations, white Gaussian noise with zero mean is first generated at the pixel level for each cube, convolved with the same beam kernel as the signal, and then scaled so its standard deviation matches the value corresponding to the chosen peak SNR, given the known maximum flux density voxel. In other words, although the added noise is drawn from a Gaussian distribution, it is not \textit{white} noise; rather, it exhibits \textit{spatial correlations} consistent with the characteristics of interferometric noise \citep{Tsukui2023}. This procedure follows established practices in interferometric imaging, as described in the \texttt{CASA} documentation \citep{casa_docs} and the ALMA Technical Handbook \citep{alma_handbook}, and ensures that the denoising problem formulation is physically realistic and aligned with the statistical properties of real ALMA data.

Visualisations of variably resolved sources, as well as variable levels of noise, are depicted in figure \ref{fig:toy_visualise}. Varying the resolutions is an interesting factor in the performance analysis of denoising methods, as central galaxies with diffuse emission after beam convolution may get diluted due to high levels of noise, and hence it would be challenging to recover the total flux and detailed morphology of the sources.

\subsection{Synthetic IFU data from cosmological simulations}\label{sec:mock_ifu}

Synthetic IFU data cubes for the Feedback in Realistic Environments (FIRE) simulation \citep{hopkins_2015, hopkins_2018, Wetzel_2023} were generated using a pipeline (\citet{Roy2025} \textit{[in prep.]}; see also \citet{richings2022}) that combines the CHIMES non-equilibrium chemistry solver \citep{Richings2014a, richings2014b}, and the RADMC-3D Monte Carlo radiative transfer code \citep{Dullemond}. The FIRE galaxy presented here is simulated from cosmological initial conditions and is part of the suite of massive galaxies \citep{feldmann, angles-alcazar, Cochrane_2024}. In particular, we focus on halo B2, which was selected to have a mass $\sim10^{13}\:M_{\odot}$ at $z=2$ \citep{feldmann}, and first presented using the FIRE-2 model in \citet{cochrane2023}.

\begin{figure}[h!]
    \includegraphics[width=0.99\columnwidth]{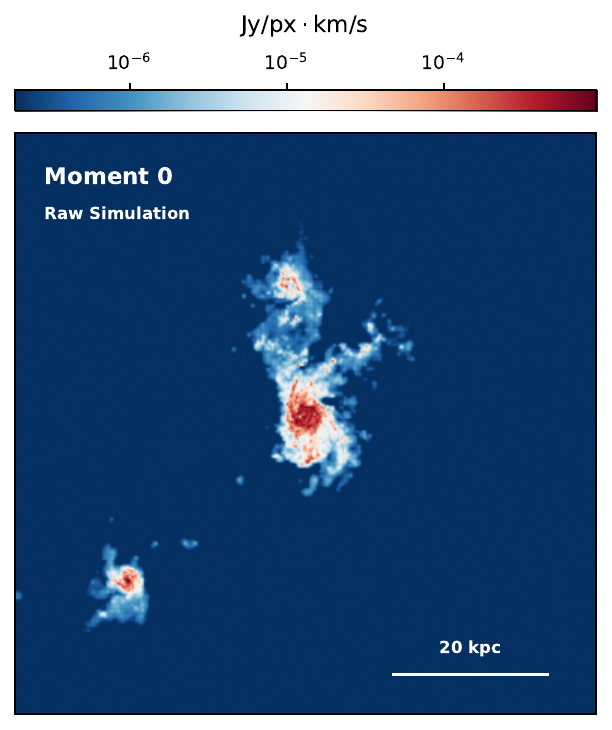}
    \includegraphics[width=\columnwidth]{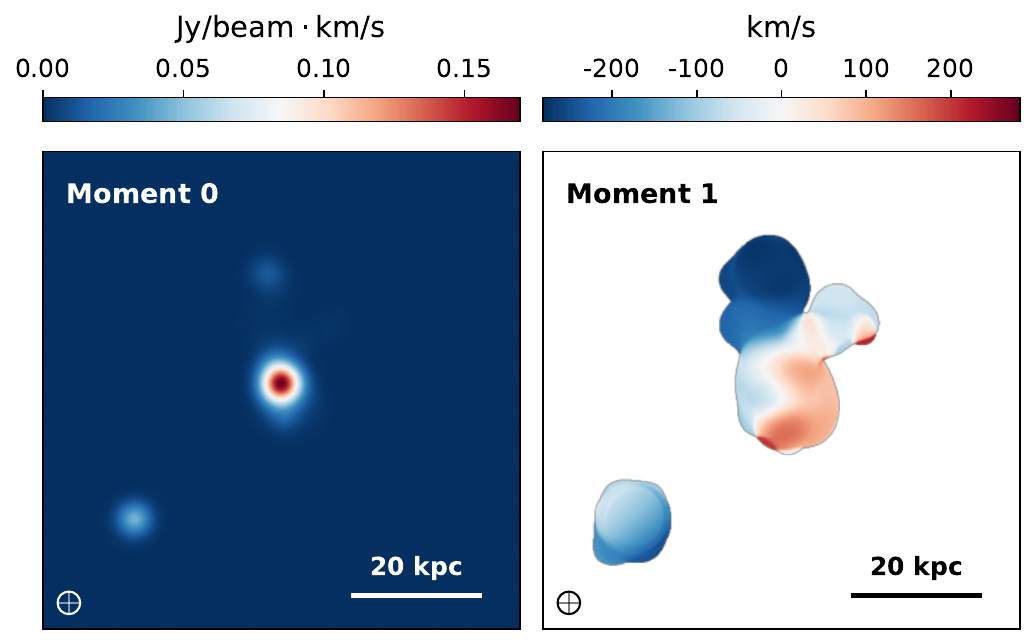}
    \includegraphics[width=\columnwidth]{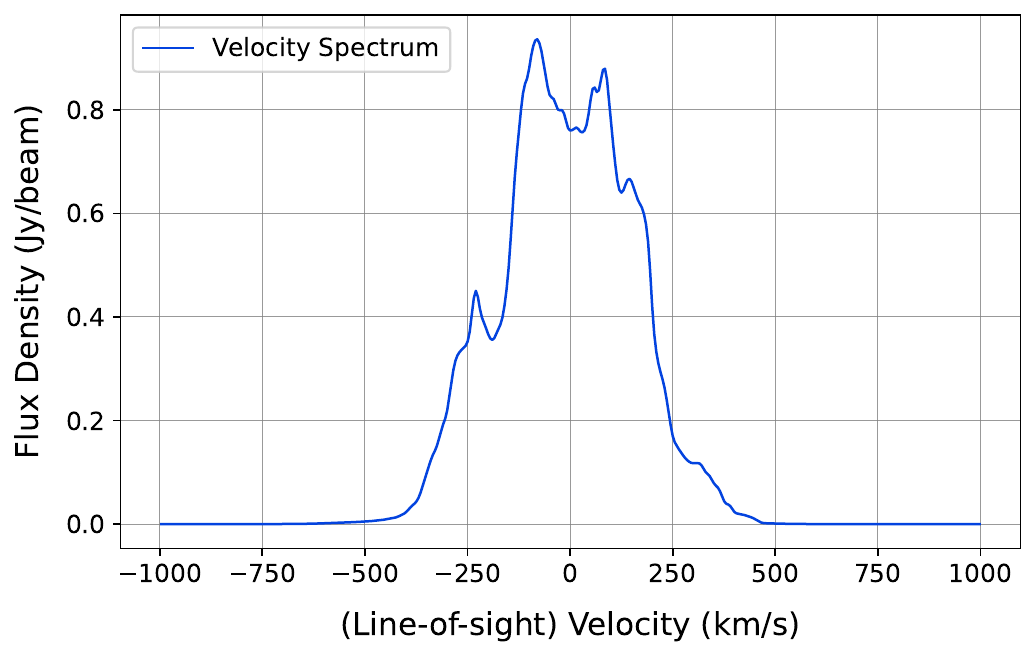}
    \caption{\textbf{Top:} Moment 0 map of the original full-resolution mock IFU from the FIRE simulation in $\rm Jy/px$; \textbf{center:} \textit{Left:} Moment 0 map after beam convolution, and \textit{Right:} Moment 1 map depicting the kinematics; \textbf{Bottom:} Line of sight velocity spectrum depicting a broad peak centred around the central velocity with multiple minor peaks.}
    \label{fig:sim_cube_preprocess}
\end{figure}

At z $\sim4.53$, this halo has a stellar mass of $0.67\times10^{11}\,\mathrm{M_{\odot}}$ within the central 26 kpc of the host galaxy and is in the process of undergoing several galaxy mergers, which represent conditions similar to those of the host galaxy of high-z multiple-merger observation - W2246–0526, hereafter referred to as W2246 (see section \ref{sec:w2246_data}) \citep{Diaz-Santos-2018}.  We generate the IFUs of B2 at z $\sim4.53$ for a 140 kpc region around the central host galaxy with a spatial resolution of 136.7 pc/pixel and spectral resolution of 5 km/s. For the purposes of this work, we zoom into a field of view of 68.35 kpc, around the central host galaxy. Specifically, we use a synthetic IFU cube of the CII emission line of this W2246 analog simulation, assuming that the level populations are dominated by collisional excitation and including the effects of dust absorption and scattering along the line of sight. We used the IFU cube of the [\ion{C}{ii}] emission line of this source in this work as a robust example of a realistic high-redshift source with a known ground truth and to draw a more direct comparison with the ALMA W2246 cube.

The mock IFU is expressed in units of specific intensity $I_{[C_{II}]}$ in the rest frame of the galaxy ($\rm erg/s/cm^2/Hz/sr$). To simulate realistic ALMA observations with instrumental effects, we preprocess the data as follows: First, we account for high-redshift cosmological effects by expressing the rest-frame specific intensity in the observer frame of view. According to Toleman's surface brightness test \citep{toleman, Weinberg1972, Peebles1993, Hogg1999Distances}, the cosmological surface brightness dimming is expressed as a factor of \((1+z)^{-4}\) when considering the bolometric (frequency-integrated) quantity. However, when focusing on the specific intensity at a fixed observed frequency band \(\nu\), an additional factor of \((1+z)\) is introduced, changing the dimming from a factor of \((1+z)^{-4}\) to \((1+z)^{-3}\):

\begin{equation}\label{eq:intensity_dimming}
\begin{aligned}
I_{\mathrm{bol}, \mathrm{obs}} &= \frac{I_{\mathrm{bol}, \mathrm{rest}}}{(1+z)^4} \\[6pt]
I_{\nu, \mathrm{obs}} &= \frac{d I_{\mathrm{bol}, \mathrm{obs}}}{d\nu} 
= \frac{d}{d\nu} \left( \frac{I_{\mathrm{bol}, \mathrm{rest}}}{(1+z)^4} \right) 
= \frac{1}{(1+z)^4} \frac{d I_{\mathrm{bol}, \mathrm{rest}}}{d\nu} \\[6pt]
&= \frac{1}{(1+z)^4} \frac{d I_{\mathrm{bol}, \mathrm{rest}}}{d\nu'} \frac{d\nu'}{d\nu} 
= \frac{1}{(1+z)^4} \frac{d I_{\mathrm{bol}, \mathrm{rest}}}{d\nu'} (1+z) \\[6pt]
&= \frac{1}{(1+z)^3} I_{\nu', \mathrm{rest}}
\quad \text{where} \quad \nu' = (1+z) \nu
\end{aligned}
\end{equation}

As we know the spatial resolution per pixel in kpc ($\ell_{px}$) and the redshift of the source, we can calculate the angle subtended by each pixel in radians ($\phi_{px}$) from the angular diameter distance ($D_A$) at the given redshift ($z$), and therefore the solid angle per pixel in steradians ($\omega_{px}$) assuming $\rm \Lambda CDM$ cosmology with the fiducial parameters of the FIRE simulations ($H_0 = 69.7 \text{ km}\mathrm{\,s^{-1}\,Mpc^{-1}}, \Omega_{\mathrm{m},0} = 0.2821. \Omega_{\Lambda,0} = 0.7179$):

\begin{equation}
\begin{aligned}
\phi_{px} &= \frac{\ell_{px}}{D_A(z)},& 
\omega_{px} = \phi_{px}^2
\end{aligned}
\end{equation}

By integrating the specific intensity cube with the solid angle per pixel, we obtain the observed flux density spectral cube and express it in units of $\rm Jy/px$. The final preprocessing step is convolving the cube with a synthesised beam - in the context of our work, the general circular beam of FWHM = 3.75 px, as in section \ref{sec:variable_dataset} - and scaling the flux density values to express the data in units of Jy/beam. A visualisation of this methodology is shown in figure \ref{fig:sim_cube_preprocess}.

\subsection{Reference Observational Dataset: \textit{ALMA-CRISTAL}}
\label{sec:cristal}

The [\ion{C}{ii}] Resolved ISM in STar-forming galaxies with ALMA (CRISTAL) survey provides an unprecedented dataset for studying galaxy evolution at $4<z<6$, a formative period covering the end of cosmic reionization. Comprising nearly 40 main-sequence star-forming galaxies, CRISTAL combines high-resolution ALMA Band 7 observations of [C\,\textsc{ii}]\,158\,$\mu$m emission and dust continuum with deep HST and JWST/NIRCam imaging for the optical and UV regimes.

The ALMA band 7 observations were carried out during Cycle 8 and Cycle 9, between December 2021 and April 2023. CRISTAL galaxies were first observed as part of the ALPINE Large Program (Le Fèvre et al. 2020). Data were processed using the CASA software (versions 5.6.1 and 6.5.2). No manual flagging was required. Individual datasets were merged into a single measurement set. The cubes used in this work for CRISTA-02 and CRISTAL-19 were generated with spectral resolutions of 10 km/s and using a "natural" weighting scheme, which maximizes the SNR and image quality of the data. The cleaning was carried out with the CASA task "tclean" using the auto-multithres algorithm (see \citet{herrera} for details on the parameters used). Continuum-subtracted cubes were created after the source continuum was identified in the line+continuum cubes and modeled using the "uvcontsub" task.

These multi-wavelength data enable spatially resolved mapping of stellar mass, star formation rate (SFR), dust content, and interstellar medium (ISM) structure on scales of $\lesssim 1$\,kpc \citep{herrera, li, ikeda}. Such resolution at these redshifts is rare, and it is critical for characterizing the physical processes governing early galaxy assembly, including gas accretion, internal feedback, and disk formation.  In this work, we use the ALMA [C\,\textsc{ii}] observations.

The CRISTAL sample spans a broad range of morphologies and dynamical states, including rotationally supported disks, compact systems, and interacting or merging galaxies \citep{lee, telikova}. Kinematic modeling reveals elevated gas velocity dispersions and turbulent disk structures. Many galaxies show extended [C\,\textsc{ii}] halos, clumpy star formation, and signatures of outflows, highlighting the complex interplay between gas dynamics, star formation, and feedback \citep{birkin, herrera}.

\begin{figure}[t!]
    \includegraphics[width=\columnwidth]{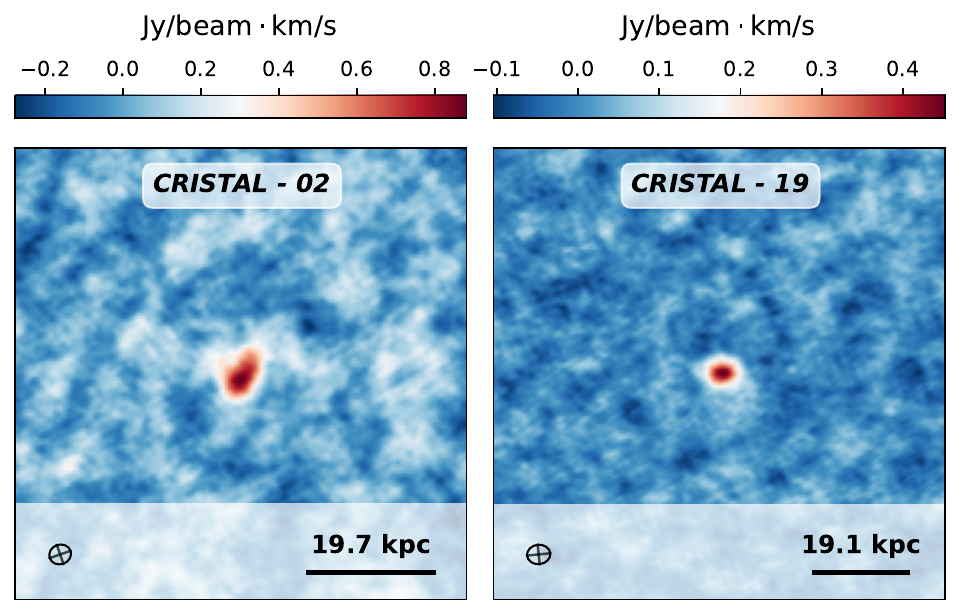}
    \includegraphics[width=\columnwidth]{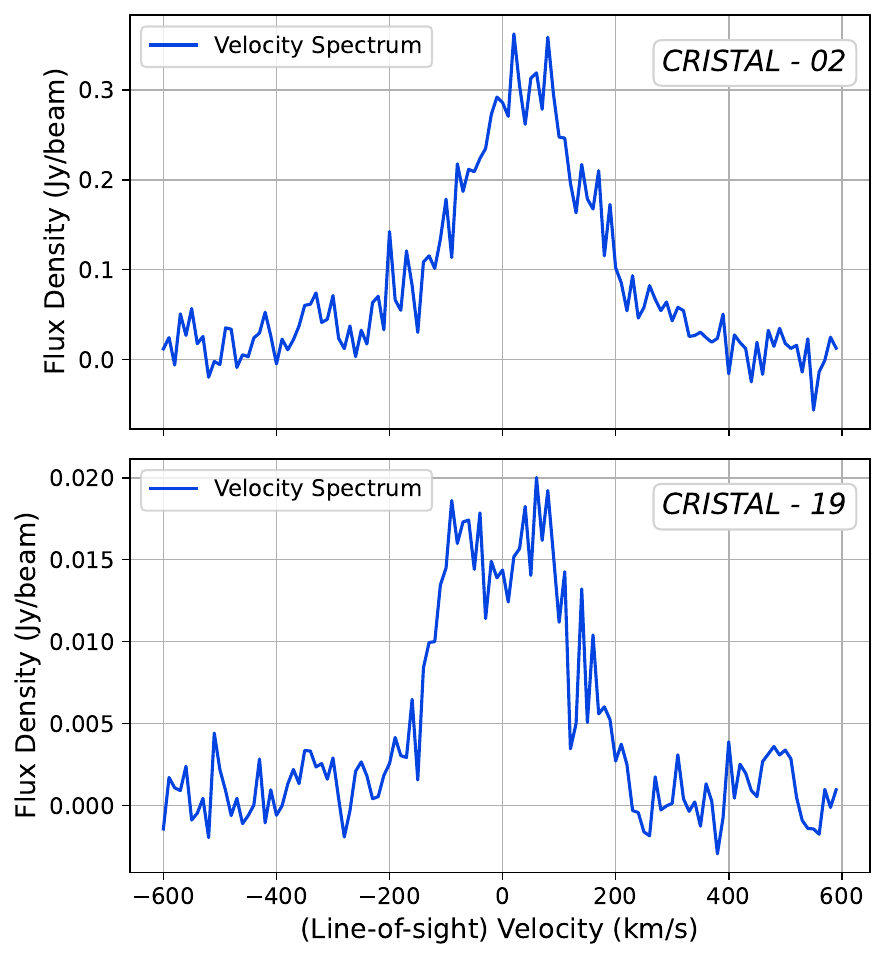}
    \caption{\textbf{Top Left:} Moment 0 map of CRISTAL-02 (resolved by the beam), and \textbf{Top Right:} CRISTAL-19 (unresolved by the beam); \textbf{Bottom:} The corresponding line-of-sight spatially integrated velocity spectra for the two spectral cubes within the emission apertures. These examples show the high levels of noise and variable resolutions explored in this survey.}
    \label{fig:cristal_vis}
\end{figure}

These features are key to understanding the baryon cycle in the early Universe, yet much of the spatial and spectral information is hampered by low signal-to-noise ratios (SNRs) and spatially correlated interferometric noise. In this context, CRISTAL serves as a benchmark dataset for the development and validation of advanced denoising algorithms. It encompasses both resolved and unresolved sources across a wide dynamic range in morphology, flux, and SNR, making it ideally suited for testing denoising performance under realistic observational conditions.

The application of such methods to CRISTAL data is expected to improve the recovery of physical quantities, including velocity fields, dynamical masses, and spatially resolved SFRs. In turn, this will enable more accurate assessments of galaxy kinematics, ISM conditions, and feedback mechanisms at the end of the reionization epoch.

In our experiments, we have used the CRISTAL-2 (z $\sim 5.3$) and CRISTAL-19 (z $\sim5.2$) observations, which are resolved and unresolved by the PSF, respectively (Figure \ref{fig:cristal_vis}).

\subsection{Complementary Observational Data: - \textit{W2246–0526}}
\label{sec:w2246_data}

\begin{figure}[t!]
    \includegraphics[width=\columnwidth]{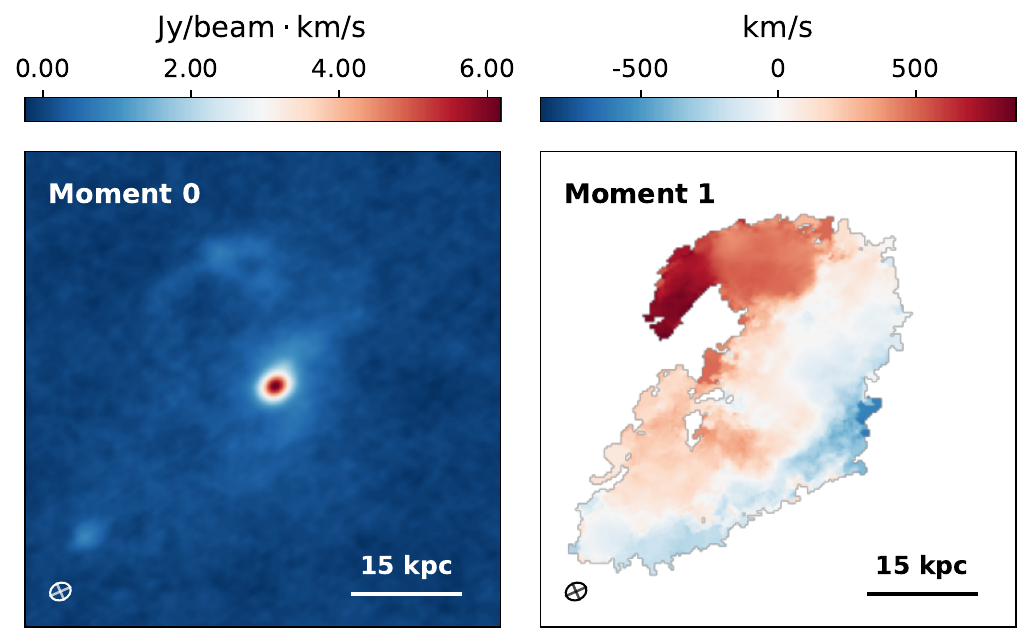}
    \includegraphics[width=\columnwidth]{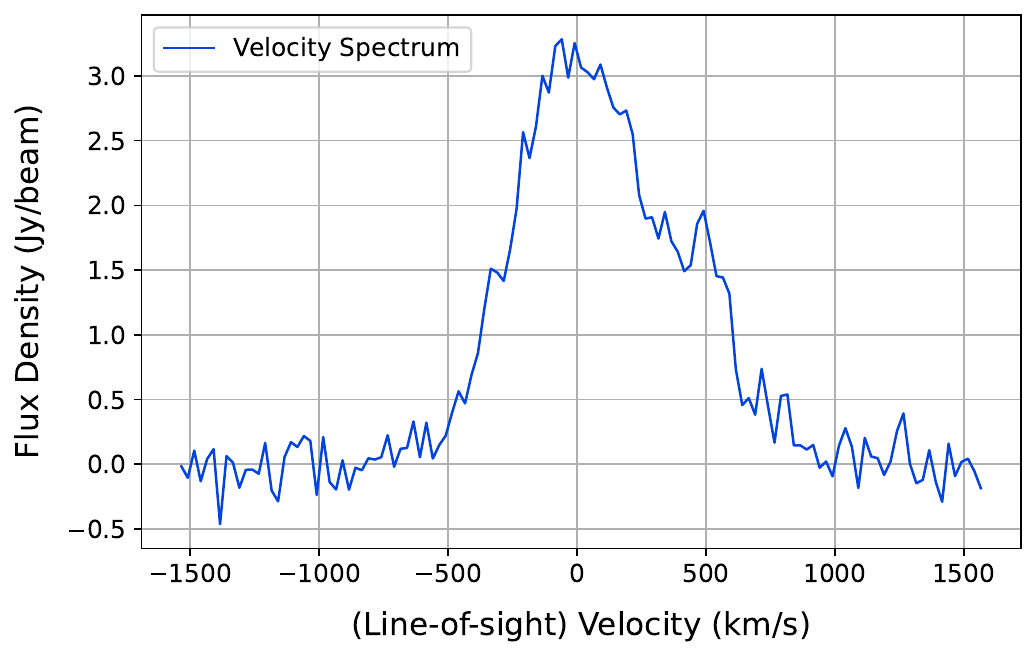}
    \caption{Spatial orientation and kinematics of \textbf{ALMA [\ion{C}{ii}] emission line} spectral cube of W2246. \textbf{Top Left:} Moment 0 map; \textbf{Top Right:} Moment 1 map within the emission region depicting the complex kinematics of the dust emission; \textbf{Bottom:} Spatially integrated line-of-sight velocity spectrum of the system within the emission region, with some noise due to instrumental effects.}
    \label{fig:moment_maps_spectrum}
\end{figure}

W2246–0526 (WISE J224607.57–052635.0), is the most luminous hot dust‐obscured galaxy (“Hot DOG”) at z $\sim4.601$ \citep{Diaz-Santos-2021}. Its bolometric luminosity is $\sim3.5\times10^{14}L_{\odot}$ \citep{Tsai_2015}, driven by a buried AGN. 

The ALMA band 7 observations for W2247 were obtained during several Cycles, between 2018 and 2022. Data were processed with the CASA software version 6.6.5. No manual flagging was required. Datasets were processed simultaneously but independently. The cube was generated with a spectral resolution of 10 km/s and using a "natural" weighing scheme, similar to the CRISTAL galaxies. The cleaning was carried out with the "tclean" task using manual masks covering the area where emission from the multiple-merger system is evident. Continuum-subtracted cubes were created after inspecting the line+continuum cube to identify line-free frequencies and model them using the "uvcontsub" task.

High‐resolution [\ion{C}{ii}] 157.7 $\mu \rm m$ observations show that the [\ion{C}{ii}] line is extremely broad ($\rm FWHM \geq 500 \text{ km}\; s^{–1}$) and essentially uniform over $\sim2.5$ kpc of the host,i.e., the ISM of W2246 is highly turbulent \citep{Diaz-Santos_2016}, with only very modest velocity gradients (no settled disk). Deep ALMA dust‐continuum imaging further reveals at least three smaller companion galaxies connected to W2246 by extended tidal “bridges” of dust spanning tens of kpc. These combined data indicate that W2246 is undergoing a multiple‐merger event: it is accreting material from its neighbors through these dusty streams while driving large‐scale outflows (rather than forming a quiescent rotating disk) \citep{Diaz-Santos-2018, roman}

Line emission such as [\ion{C}{ii}] can be faint, spatially extended, and embedded within strong thermal dust continuum \citep{Diaz-Santos-2021}. High-redshift ALMA observations are often sensitivity-limited and subject to high levels of instrumental noise, particularly in faint, diffuse emission regions. These factors lead to a low signal-to-noise ratio (SNR) in many spectral channels and spaxels.

Consequently, advanced denoising techniques are potentially extremely useful for recovering astrophysical signals, improving the fidelity of flux and kinematic measurements, and enabling reliable morphological analysis of both the host galaxy and its extended structures.

We include W2246 as an additional observational test case due to its characteristic diffuse emission and dynamically disturbed, non-rotating morphology. For our supervised denoising method (\ref{sec:supervised}), while our synthetic training and validation data, as well as the CRISTAL survey references are based on rotating galaxies without diffuse components, the model learns general spectral and spatial patterns. Testing on W2246 thus offers an opportunity to evaluate the robustness and limitations of our denoising approach under markedly different physical conditions.

\section{Evaluation criteria}\label{sec:evaluation}

\subsection{Identification of Emission regions}

As we aim to satisfactorily denoise data while preserving the \textit{emission flux} to the fullest possible extent, a major step is identifying the emission regions to create an aperture to perform statistical analyses.

In the absence of a true noise-free reference cube, as is invariably the case for astronomical observations, the quantitative assessment of denoising performance necessitates a well-defined and physically motivated prescription for selecting emission regions. To enable method-to-method comparability, we restrict our analysis to the emission associated with the central galaxy and employ a fixed, circular aperture centred on its coordinates and applied uniformly to all spectral channels of each cube. As defined in the sample design of the toy models (Section \ref{subsection:toy}), \(n \geq 0\) satellite galaxies are included to introduce realistic confusion and flux mixing that emulate galaxy interactions similar to our simulated and observational references.

In general, astronomical sources that have already been observed in the past can provide us with a prior on their spatial extent. This prior can be used to estimate their expected size at other wavelengths, especially when direct measurements are not available. As a conservative and observationally grounded choice, we define the aperture diameter $D_{\text{ap}}$ according to the following criterion:

\begin{equation}
D_{\text{ap}} =
\begin{cases}
2 \times D_e & \text{if } D_e > \text{FWHM}_{\text{beam}} \\
2 \times \text{FWHM}_{\text{beam}} & \text{if } D_e \leq \text{FWHM}_{\text{beam}} \text{ or if } D_e \text{ is unknown}
\end{cases}
\end{equation}

\begin{figure}[t!]
    \includegraphics[width=\columnwidth]{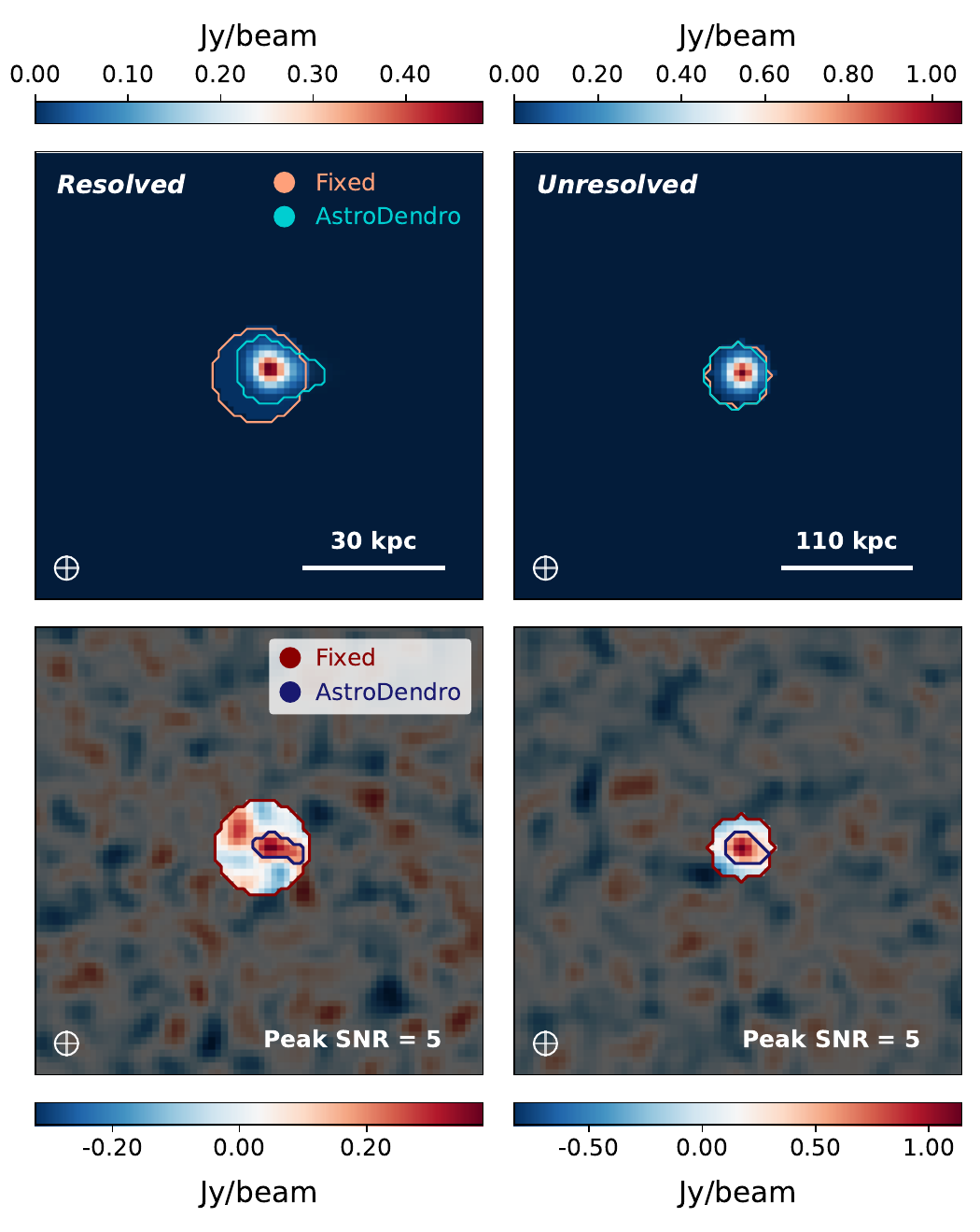}
    \caption{\textbf{Top:} slices of two noise-free cubes with varying beam resolutions and the respective fixed and dendrogram-based apertures, depicting how the fixed aperture gets scaled with respect to the beam size; \textbf{Bottom:} The same slices after application of spatially correlated Gaussian noise with peak SNR = 5, depicting the fixed aperture and how the dendrogram-based apertures are not able to identify the complete region of flux due to the high level of noise. This also depicts the expansion of the initial aperture for larger beam sizes.}
    \label{fig:emission_masks}
\end{figure}

Here, $D_e = 2R_e$ is the effective diameter of the source based on prior knowledge from the input simulation parameters. The beam is circular, with a full width at half maximum $\text{FWHM}_{\text{beam}}$ expressed in pixels.

This choice ensures full coverage of extended emission when the source is resolved (i.e., $D_e > \text{FWHM}_{\text{beam}}$), and falls back to a conservative beam-based aperture when the source is unresolved or when its size is not known \textit{a priori}. This approach reflects common observational strategies and provides a robust, realistic framework for evaluating denoising performance across a range of source sizes and noise levels.

We draw a direct comparison of our fixed apertures with the known ground truth for verification of the feasibility, by identifying the precise emission aperture using $n$-dimensional dendrograms from \texttt{AstroDendro} to hierarchically extract structures from the ground truth.

Figure \ref{fig:emission_masks} serves as a visual reference where the top row is one slice of the clean ground truth of resolved and unresolved toy central galaxies, respectively; and the bottom row shows the corresponding noisy slices. It is evident that while dendrograms can identify the emission regions with great accuracy in the ground truth, this method does not suit the cubes with high levels of noise that we aim to analyze in this paper. It is also evident that the approximately estimated circular aperture constructed from no knowledge of the ground truth can contain the entire emission region in the ground truth with marginal error, and can be used on the noisy maps.

\subsection{Performance metrics}

Our primary goal is to conserve the total flux within the emission aperture identified in the previous section while evaluating denoising performance quantitatively. To this end, we consider the following metrics:

\subsubsection{Total flux conservation}  
We measure the total flux within the emission aperture for both the denoised and ground-truth cubes. Let $X_{i,j,k}^{\text{den}}$ and $X_{i,j,k}^{\text{true}}$ represent the voxel flux densities of the denoised and ground truth cubes for corresponding noisy flux densities $Y_{i,j,k}$, and let $\mathcal{A}$ denote the set of voxels inside the emission aperture. The total fluxes are:

\begin{equation}
\begin{split}
S_{\text{den}} &= \sum_{(i,j,k) \in \mathcal{A}} X_{i,j,k}^{\text{den}} \:\:,\:
S_{\text{true}} = \sum_{(i,j,k) \in \mathcal{A}} X_{i,j,k}^{\text{true}}
\end{split}
\end{equation}

We aim to preserve the flux such that $S_{\text{den}}$ is as close as possible to $S_{\text{true}}$.

\subsection{Root Mean Squared Error (RMSE)}  
To quantify the fidelity of the denoised signal, we calculate the RMSE between the denoised and ground truth flux within the emission aperture:

\begin{equation}
\mathrm{RMSE}_{\text{ap}} = \sqrt{ \frac{1}{N_p} \sum_{(i,j,k) \in \mathcal{A}} \left( X_{i,j,k}^{\text{den}} - X_{i,j,k}^{\text{true}} \right)^2 }
\end{equation}

where $N_p = |\mathcal{A}|$ is the number of voxels in the aperture.

\vspace{2mm}

Together, these metrics allow us to assess whether the denoising method effectively conserves total flux, accurately reconstructs the emission structure within the aperture, and minimizes residual artifacts. We note that these metrics are specifically chosen to quantify performance for individual reconstructions (point estimates), rather than assessing differences between distributions of samples. Since the methodologies introduced in Sections \ref{sec:unsupervised} and \ref{sec:supervised} provide point-estimate reconstructions, this approach aligns with our specific aim of evaluating deterministic recovery fidelity. Future work attempting uncertainty quantification, however, would require metrics specifically designed to handle uncertainty calibration for high-dimensional inference.

\section{Unsupervised Methods} \label{sec:unsupervised}

In this section, we describe the \textit{unsupervised} denoising strategies explored in this study. These methods do not require any labeled training data or prior knowledge of the clean signal. Instead, they exploit intrinsic statistical, structural, or sparsity properties of the data to separate signal from noise. We evaluate three representative approaches: Principal Component Analysis (PCA), which leverages linear correlations; Independent Component Analysis (ICA), which assumes statistical independence; and iterative 2D-1D wavelet thresholding, which takes advantage of multiscale sparsity. Each of these methods provide useful benchmarks for the denoising task.

\subsection{PCA-Based Denoising}

Principal Component Analysis (PCA) \citep{Hotelling_PCA, Jolliffe_PCA, Pearson_PCA} is a widely used statistical technique for dimensionality reduction that can also serve as an effective, physically motivated, and statistically principled method for denoising spectral data cubes.

For a data cube $\bm{Y}$ with spatial dimensions $n_x \times n_y$ and spectral dimension $n_s$, we reshape it into a two-dimensional data matrix $\mathbf{Y'} \in \mathbb{R}^{(n_x n_y) \times n_s}$, where each row corresponds to the full spectrum of a spaxel, and each column corresponds to flux values across all spaxels for a given spectral channel. Prior to PCA, the matrix is mean-centred along each spectral channel by subtracting the column-wise mean, ensuring zero mean per feature.

PCA decomposes the centred matrix into orthogonal principal components (PCs), which are linear combinations of the original spectral channels. Each PC corresponds to an eigenvector of the covariance matrix, and the associated eigenvalue quantifies the variance captured by that component. PCs are sorted in descending order of eigenvalues, so that the first PC captures the largest variance, the second captures the next largest variance orthogonal to the first, and so forth.

\subsubsection{Denoising Procedure}

To denoise the cube, we iteratively reconstruct it using an increasing number of PCs, starting with $k=2$ and incrementing by one at each step. The reconstructed cube at each iteration includes only the first $k$ PCs, capturing the dominant signal features, while the remaining components are treated as noise and discarded.

The optimal number of PCs, $n_{\text{PC}}$, is guided by flux conservation within the previously defined emission aperture. Specifically, $n_{\text{PC}}$ is chosen when the total flux within the aperture stabilizes, indicated by a plateau in flux over three consecutive increments of $k$. This plateau shows that additional components no longer significantly alter the reconstructed signal, avoiding the inclusion of noise-dominated PCs.

If the flux plateau criterion is not met, for example, due to subtle or complex emission structures, the iteration is terminated when the \textit{cumulative explained variance} exceeds 70\%. The cumulative explained variance up to the $k$-th PC measures the fraction of total variance captured by the first $k$ components, and is defined as
\begin{equation}
\mathrm{Cumulative\ Explained\ Variance}(k) = \frac{\sum_{i=1}^k \lambda_i}{\sum_{j=1}^p \lambda_j},
\end{equation}
where $\lambda_i$ is the eigenvalue of the $i$-th PC and $p$ is the total number of PCs. This threshold ensures that a sufficiently large fraction of the total variance, and hence the underlying signal, is captured without overfitting. Selecting $n_{\text{PC}}$ at this point acts as a robust fallback, preventing noise-dominated reconstructions.

\subsection{ICA-Based Denoising}

Independent Component Analysis (ICA) \citep{Comon1994, Hyvarinen2001} is a blind source separation technique that decomposes multivariate data into a linear mixture of statistically independent components. Unlike PCA, which maximizes variance and yields uncorrelated components, ICA seeks a transformation that maximizes statistical independence, typically measured via non-Gaussianity (e.g., kurtosis or negentropy) or minimized mutual information.

For spectral cube denoising, the data cube $\bm{Y}$ is reshaped into a two-dimensional matrix $\mathbf{Y'} \in \mathbb{R}^{(n_x n_y) \times n_s}$, where each row corresponds to a spaxel's spectrum. The ICA algorithm estimates a mixing matrix $\mathbf{A}$ and a source matrix $\mathbf{S}$ such that $\mathbf{Y'} \approx \mathbf{A} \mathbf{S}$, with the rows of $\mathbf{S}$ representing statistically independent components.

\subsubsection{Denoising Procedure}

ICA assumes that the observed data is a linear combination of latent sources, where the structured, coherent signal is contained in a small number of strongly independent components, while the noise, being random and diffuse, contributes weakly across many components. This separation enables implicit denoising: reconstructing the cube using only the dominant components preserves signal while suppressing noise.

To determine the optimal number of independent components, $n_{\text{IC}}$, we iteratively reconstruct the cube with an increasing number of components. At each iteration, the reconstructed data is evaluated using a flux-based metric, and $n_{\text{IC}}$ is chosen when the total flux within the emission aperture plateaus over three successive increments, indicating convergence.

If the flux plateau criterion is not met, for instance due to the absence of a clear plateau, a fallback strategy is used: $n_{\text{IC}}$ is selected as the number of components minimizing the residual noise, defined as the standard deviation of the residual between the noisy input and the reconstructed data within the unmasked, signal-containing region. This assumes better denoising corresponds to minimizing residual fluctuations, though it does not prioritize flux conservation.

\begin{figure*}[h!]
    \includegraphics[width=1\textwidth]{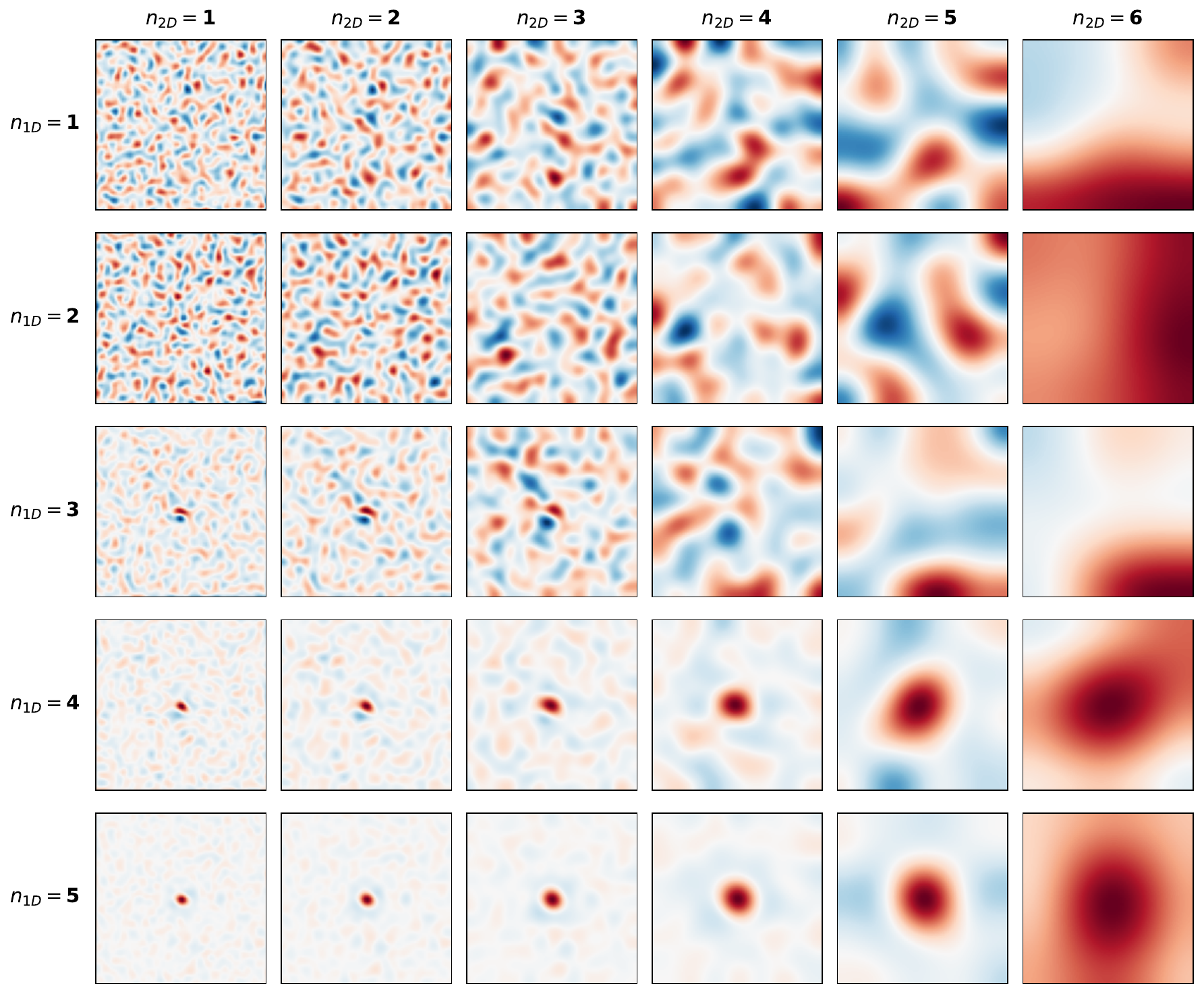}
    \caption{Visualisation of one 2D slice of every 2D-1D scale for the wavelet decomposition of a single noisy spectral cube with one central galaxy situated at the center of the FoV. The wavelet coefficients corresponding to the signal can be visually identified in the larger (low frequency) scales. High frequency wavelet coefficients (low $n_{1D} \text{ and } n_{2d}$) correspond to noise. The last scale ($n_{2D}=6, n_{1D}=5)$ corresponds to the coarse scale/approximation coefficients containing information about the average flux.}
    \label{fig:2d1d_trans}
\end{figure*}

\subsection{2D-1D Wavelet-Based Denoising}

Wavelet transforms and sparsity-enhancing techniques have been prominent in data analysis in the astronomical context \citep{Starck2006, Mallat1999}. In this methodology, we adopt a hybrid wavelet decomposition scheme tailored to the structure of spectral cubes - applying a 2D isotropic wavelet transform to each spatial slice and a 1D wavelet transform along the spectral axis for each spaxel.

This 2D-1D approach, previously used by \citet{gamma_LAT_2d1d} and \citet{2d1d_source_finding} for source identification, exploits the anisotropic nature of the data: spatial correlations are isotropic across the sky (justifying a 2D isotropic transform), whereas spectral correlations evolve independently and require a dedicated 1D analysis. A fully isotropic 3D wavelet transform fails to capture such directional differences and can dilute spectral features, motivating our separable decomposition.

Let $\bm{X}$ denote a spectral cube. The forward wavelet transform is written as:
\begin{equation}
    \bm{\alpha} = \bm{\Phi}^T \bm{X}
\end{equation}
where $\bm{\alpha}$ is the set of 2D-1D wavelet coefficients and $\bm{\Phi}^T$ is the decomposition operator.

In our case, the 2D transform is the undecimated isotropic Starlet transform \citep{starlet}, a non-orthogonal wavelet well-suited for preserving extended structures in astronomical images while maintaining spatial resolution. The 1D transform is a decimated biorthogonal wavelet transform using the B-spline wavelet of order (4,4), part of the Cohen–Daubechies–Feauveau (CDF) family \citep{cdf}, equivalent to the CDF-9/7 wavelet. Figure \ref{fig:2d1d_trans} shows the decomposition of a single slice of different scale representations of a spectral cube with one galaxy at the centre of the field-of-view.

Therefore, $\bm{\Phi}^T \neq \bm{\Phi}^{-1}$. The signal can be approximately reconstructed as:
\begin{equation}
    \bm{X}' = \bm{\Phi} \bm{\alpha}
\end{equation}

Given our noisy data $\bm{Y} = \bm{X}^{\rm true} + \bm{N}$, the denoising optimization problem can be formulated as:

\begin{equation}
J = \|\bm{Y} - \bm{X} \|^2 + \lambda \| \bm{\Phi}^T \bm{X} \|_p
\end{equation}

The first term represents the \textit{data fidelity}, ensuring that the denoised output remains close to the input data, while the second term is a \textit{regularisation} term that promotes sparsity. Here, $\lambda$ denotes the regularisation or thresholding parameter, and $p \in \{0, 1\}$ corresponds to hard or soft thresholding, respectively (\ref{sec:thresholding}). The denoised signal is then given by:
\begin{equation}
    \bm{X}^{\rm den} = \bm{\Phi} \Bigl( \delta_\lambda ( \bm{\Phi}^T \bm{Y}) \Bigr)
\end{equation}

where $\delta_\lambda$ is the non-linear thresholding operator applied to the wavelet coefficients.

\subsubsection{Thresholding} \label{sec:thresholding}

Thresholding plays a central role in wavelet-based denoising techniques. The primary objective is to suppress noise while preserving the salient features of the underlying signal. Wavelet decomposition yields a sparse multiscale representation in which noise is predominantly observed in the high-frequency detail coefficients, while signal structures tend to be concentrated in a few significant coefficients across scales. By applying an appropriate threshold to the wavelet coefficients, one can selectively attenuate or eliminate noise-dominated components while retaining meaningful features.

Importantly, the thresholding is applied to all wavelet detail scales except the coarse (approximation) scale. The coarse scale captures the large-scale, low-frequency structure of the data and contains the non-zero mean component of the signal. Since the coarse scale is not zero-mean, thresholding it would risk suppressing genuine low-frequency emission, leading to a loss of total flux or structural integrity. In contrast, the detail coefficients at each scale are, by construction, zero-mean and are thus amenable to thresholding without affecting the overall flux normalization.

The threshold applied at each detail scale depends on the noise level within that scale. A robust estimate of the noise standard deviation, $\sigma_{\alpha}$ is obtained using the \textit{Median Absolute Deviation} (MAD) \citep{gauss1816, walker1931} of the wavelet coefficients:
\begin{equation}
\label{eq:mad}
\sigma_{\alpha} = \frac{\mathrm{MAD}}{0.6745}, \quad \text{where} \quad \mathrm{MAD} = \mathrm{median}(|\alpha_i - \mathrm{median}(\alpha)|).
\end{equation}

For pure white Gaussian noise of standard deviation $\sigma_D$, the threshold at each 2D-1D scale ($j_1, j_2$) could be derived directly from $\sigma_D$. In our noise model, the noises is spatially correlated (i.e., the noise fluctuates at the beam-level, not the pixel-level), and the MAD approach represents a good alternative. MAD does not assume uncorrelated (white) noise, remains robust to outliers introduced by the signal, and provides an empirical and reliable estimate of the underlying noise fluctuations at each scale. The scaling factor $0.6745^{-1}$ ensures consistency with the standard deviation.

Given a thresholding parameter $\lambda$, the effective threshold for each coefficient becomes $\lambda\sigma_{\alpha}$. Two widely used thresholding schemes are \textit{hard thresholding} and \textit{soft thresholding}.

\textbf{Hard thresholding} sets wavelet coefficients with absolute value below the threshold to zero while preserving the others:

\begin{equation}
\mathcal{T}_{\text{hard}}(\alpha_i, \lambda) =
\begin{cases}
\alpha_i, & \text{if } |\alpha_i| \geq \lambda\sigma_{\alpha}, \\
0, & \text{otherwise}.
\end{cases}
\label{eq:hard_thresh}
\end{equation}
This method enforces sparsity by eliminating small coefficients entirely, akin to $\ell_0$-norm regularization. While it retains large coefficients without alteration, the discontinuity at the threshold ($\alpha = \pm \lambda\sigma_{\alpha}$) can introduce artifacts and instabilities in the reconstructed signal.

In contrast, \textbf{soft thresholding} shrinks all coefficients toward zero by the threshold amount, resulting in a continuous and differentiable function:
\begin{equation}
\mathcal{T}_{\text{soft}}(\alpha, \lambda) = \text{sign}(\alpha) \cdot \max\left(|\alpha| - \lambda\sigma_{\alpha}, 0\right),
\end{equation}
where the sign function is defined as
\begin{equation}
\text{sign}(\alpha) =
\begin{cases}
1, & \text{if } \alpha > 0, \\
0, & \text{if } \alpha = 0, \\
-1, & \text{if } \alpha < 0.
\end{cases}
\end{equation}
It is the proximal operator of the $\ell_1$-norm, and avoids discontinuities and reduces the risk of introducing reconstruction artifacts. However, it introduces a significant bias by shrinking even the significant coefficients, which may attenuate the signal if the threshold is too aggressive \citep{Donoho1994}.

\begin{figure*}[t!]

    \includegraphics[width=1\textwidth]{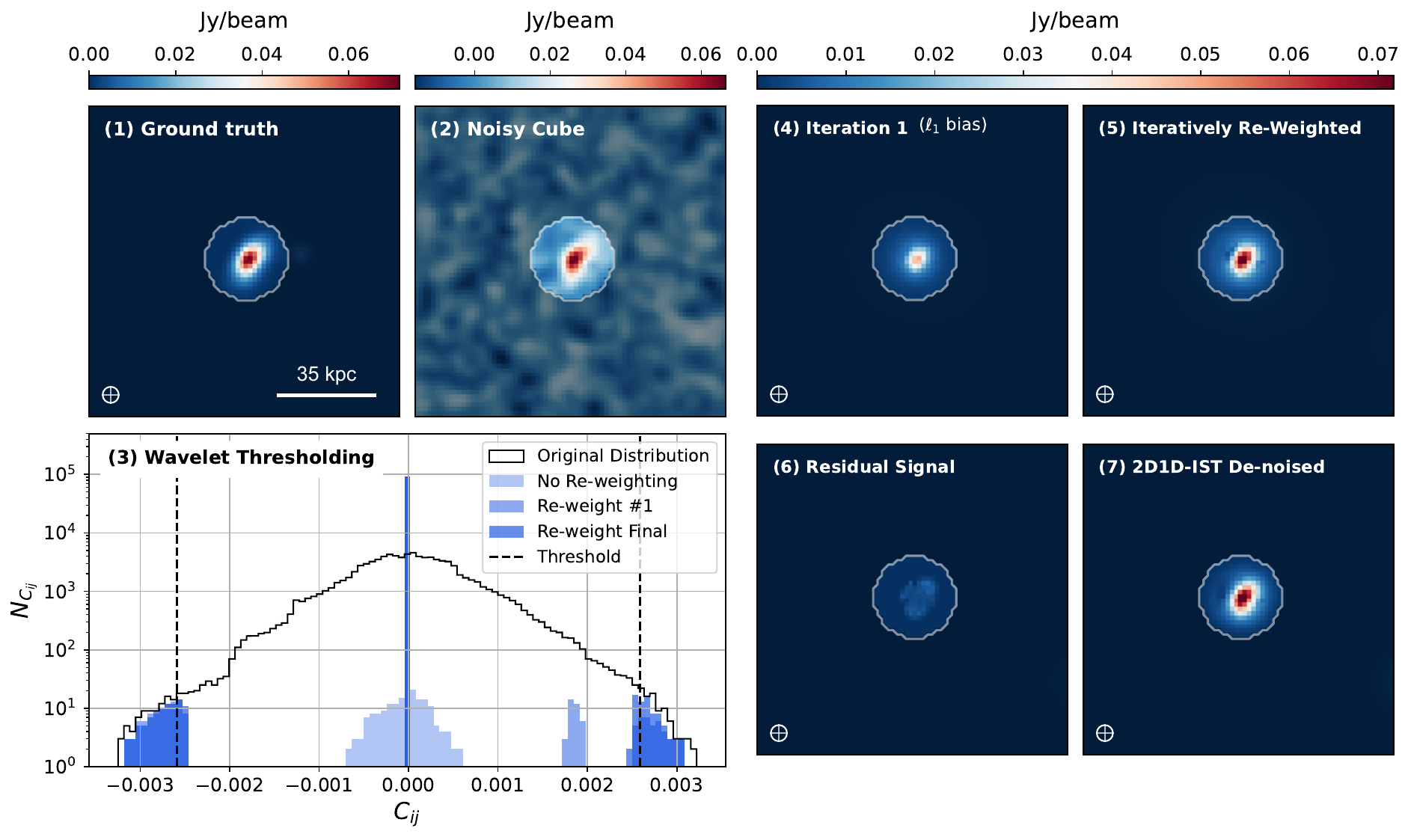}
    \caption{Step-wise depiction of iterative re-weighted 2D-1D soft thresholding-based denoising. Sub-figures \textbf{(1)} and \textbf{(2)} display the ground truth and the noisy cube with a peak SNR $\sim15$, respectively. The histogram \textbf{(3)} illustrates the distribution of wavelet coefficients at a single 2D-1D scale, as well as the thresholded coefficients at consecutive reweighting iterations. Sub-figures (4) through (7) highlight the results at different stages of the denoising process: \textbf{(4)} single-step soft thresholding with $l_1$ biased coefficient amplitudes, \textbf{(5)} de-biased result of iteratively reweighting the wavelet coefficients, \textbf{(6)} the total signal remaining in the residual that can be extracted by iteratively thresholding the residuals, and \textbf{(7)} Final de-biased and de-noised output.}
    \label{fig:IST_steps}
\end{figure*}

The choice between hard and soft thresholding depends on the trade-off between preserving signal amplitude and avoiding reconstruction artifacts. Hard thresholding retains large coefficients exactly but can produce discontinuities. Soft thresholding offers greater numerical stability and smoother results but can reduce the amplitude of true signal components. In practice, adaptive or hybrid approaches are often used to combine the benefits of both methods.

\subsubsection{Iterative Soft Thresholding}

As the simple soft thresholding method results in a systematic underestimation of true signal amplitudes, we aim to correct for this $\ell_1$ bias while preserving the convex and differentiable nature of soft thresholding, by implementing an \textit{iterative reweighting strategy}. Furthermore, upon the application of conservative cuts, some signal might be left in the residual, hence, we incorporate a secondary residual de-biasing/signal extraction step to extract the maximum possible signal iteratively.\\

\noindent\textbf{Re-weighting:} The process begins with a unweighted single-step soft thresholding. From the second iteration onward, weights are computed dynamically and iteratively from the denoised wavelet coefficients of the previous iteration to minimize the bias introduced in the first step. For each coefficient in the $n^{\rm th}$ iteration, the corresponding weight is defined as:

\begin{equation}
\mathcal{W}^{(n)}_{ij} =
\begin{cases}
\dfrac{\lambda\sigma_{\alpha^{(n-1)}}}{|\alpha^{(n-1)}_{ij}| + \epsilon}, & \text{if } |\alpha^{(n)}_{ij}| \geq \lambda \sigma^{(n)}_{\alpha} \\
1, & \text{otherwise}
\end{cases}
\end{equation}

\noindent where:
\begin{itemize}
    \item $n \geq 2$ is the iteration index,
    \item $\alpha^{(n)}_{ij}$ are the wavelet coefficients after decomposition in iteration\textit{n}.
    \item $\sigma_{\alpha^{(n)}}$ is the estimated noise standard deviation at the corresponding wavelet scale, computed via the Median Absolute Deviation (MAD) at iteration \textit{n},
    \item $\epsilon$ is a small positive constant to avoid division by zero.
\end{itemize}

This reweighting scheme reduces penalization on high-amplitude coefficients (likely signal) while still shrinking low-amplitude coefficients (likely noise), thus retaining the smooth thresholding nature of $\ell_1$ regularization while mitigating the bias.

At each iteration, this reweighted soft-thresholding is applied to the input data $Y$, yielding the updated denoised output $X^{(n)}$. The optimization problem in analysis form is:

\begin{align}
\mathbf{X}_{\mathrm{den}}^{(n)} 
&= 
\operatorname{prox}_{\iota_{\mathbb{R}_{+}}} 
\Bigl(
\mathbf{\Phi} \, 
\mathcal{T}_{\lambda, \mathcal{W}^{(n-1)}} 
\bigl(\mathbf{\Phi}^{\top} \mathbf{Y}\bigr)
\Bigr) \\[2mm]
&= 
\max \Bigl[
\mathbf{\Phi} \, 
\mathcal{T}_{\lambda, \mathcal{W}^{(n-1)}} 
\bigl(\mathbf{\Phi}^{\top} \mathbf{Y}\bigr),\, 0
\Bigr].
\end{align}

\noindent where $\mathcal{T}_{\lambda, W}$ denotes weighted soft-thresholding applied to the wavelet coefficients, and $\operatorname{prox}_{\iota_{\mathbb{R}_+}}$ is the proximal operator of the indicator function of non-negative numbers, enforcing a positivity constraint on the denoised output in each iteration.

Once the weights converge - defined by the standard deviation of the residual between the current output and the noisy input,

\[
\bm{R^{(n)}} = \bm{Y} - \bm{X^{(n)}}_{\rm den},
\]

\noindent reaching a plateau, i.e., when $\sigma_{\bm{R^{(n)}}}$ does not decrease significantly across consecutive iterations, the de-biasing step is initiated.\\

\noindent\textbf{Residual Signal Extraction:} The residual after the final re-weighting step is wavelet-transformed, and the final weights, learned upon convergence of the re-weighting step, are applied to perform weighted soft thresholding on the residual coefficients. The resulting denoised residual is reconstructed and added to the previous output:

\begin{equation}
\bm{X}^{(n+1)}_{\rm den} = 
\operatorname{prox}_{\iota_{\mathbb{R}_+}} \Biggl(
\bm{X}^{(n)}_{\rm den} + 
\bm{\Phi} \Bigl(
\mathcal{T}_{\lambda, \mathcal{W}^{(\mathrm{final})}} 
\bigl( \bm{\Phi}^T (\bm{Y} - \bm{X}^{(n)}_{\rm den}) \bigr)
\Bigr)
\Biggr)
\end{equation}

The stopping criterion for this debiasing step is also based on plateauing minimisation of the residual standard deviation, ensuring that iterative updates are performed only as long as meaningful signal recovery occurs. Figure \ref{fig:IST_steps} depicts the step-wise denoising process.

\section{Supervised Methods} \label{sec:supervised}

After examining unsupervised approaches, we now turn to \textit{supervised} denoising techniques, which learn an explicit mapping from noisy to clean data using labeled examples. Such methods, particularly those based on deep neural networks, can model complex and highly non-linear relationships in the data that are difficult to capture with classical techniques. Motivated by this, we employ a three-dimensional U-Net, an encoder-decoder architecture with skip connections that simultaneously captures large-scale context and preserves fine spatial information, making it especially effective for denoising high-dimensional astronomical data. We train this model on our synthetic dataset of galaxy spectral cubes (Section~\ref{sec:variable_dataset}) to suppress noise while preserving astrophysical signals with high fidelity. The architecture, training procedure, and evaluation of the supervised method are detailed below.

\subsection{Deep Learning: U-Net-Based Denoising} \label{sec:unet_denoising}

To perform supervised denoising of three-dimensional (3D) spectral cubes, we adopt a deep convolutional neural network based on the U-Net architecture \citep{ronneberger2015u}, extended to three dimensions. This network is trained on the large set of synthetic toy spectral cubes (Section~\ref{sec:variable_dataset}), carefully constructed to span a wide range of spatial resolutions, morphologies, and noise conditions representative of realistic observational scenarios. The primary objective is to suppress noise while faithfully preserving both the total flux and the spatial–spectral morphology of the embedded astrophysical sources.

The U-Net follows an encoder–decoder paradigm with skip connections that link corresponding levels of the encoder and decoder paths. This design enables the network to capture both global context and fine-scale details, which is particularly crucial for denoising tasks where faint, extended structures coexist with compact sources. The architecture comprises four hierarchical levels in each path. Each encoder block consists of two 3D convolutional layers with kernel size $3 \times 3 \times 3$, without biases, followed by a downsampling operation. To mitigate boundary artifacts and ensure continuity of the signal near the cube edges, we employ reflective padding before each convolution. Downsampling is implemented via average pooling with a stride of two, which preserves the integrated flux by evenly distributing intensity, in contrast to maximum pooling or strided convolutions that may bias the signal toward high-intensity regions. Non-linearity is introduced through a \texttt{LeakyReLU} activation function applied after each convolution.

At the network’s bottleneck lies a latent representation encoded by two additional 3D convolutional layers. This central block learns compressed feature representations that capture complex spatial and spectral correlations within the cube, including subtle kinematic or morphological patterns.

The decoder path mirrors the encoder structure, progressively reconstructing the cube at higher spatial–spectral resolution. Each decoding block starts with a 3D transposed convolution (kernel size $2 \times 2 \times 2$) to upsample the feature maps. These upsampled maps are concatenated with the corresponding encoder features via skip connections, effectively reintroducing fine-scale information that may have been lost during downsampling. This is followed by two more 3D convolutional layers with the same configuration as in the encoder. A schematic illustration of the full architecture is provided in figure \ref{fig:visualise_unet}.

For training, we generate a dataset of 20,000 synthetic cubes, each with uniquely varying parameters 
(Section \ref{sec:variable_dataset}). This dataset is randomly partitioned into 80\% for training, 10\% for validation, and 10\% for testing. The network is trained with a batch size of 32, and optimized using the Adam optimiser \citep{adam_opt} with a learning rate of $10^{-4}$. Training is performed on a single NVIDIA A100 SXM4 GPU with 40 GB of memory. 

The loss function is the mean squared error (MSE), which ensures local fidelity but does not explicitly enforce global morphological accuracy. This is mitigated by our incorporation of architectural features such as average pooling and skip connections, which encourage the propagation of multiscale features and help preserve both global flux and fine structural details. Moreover, the physically motivated and morphologically diverse training dataset acts as an implicit regularizer, guiding the network toward realistic, flux-conserving reconstructions.

\section{Results} \label{sec:results}

\begin{figure*}[t!]
    \centering
        \includegraphics[width=\textwidth]{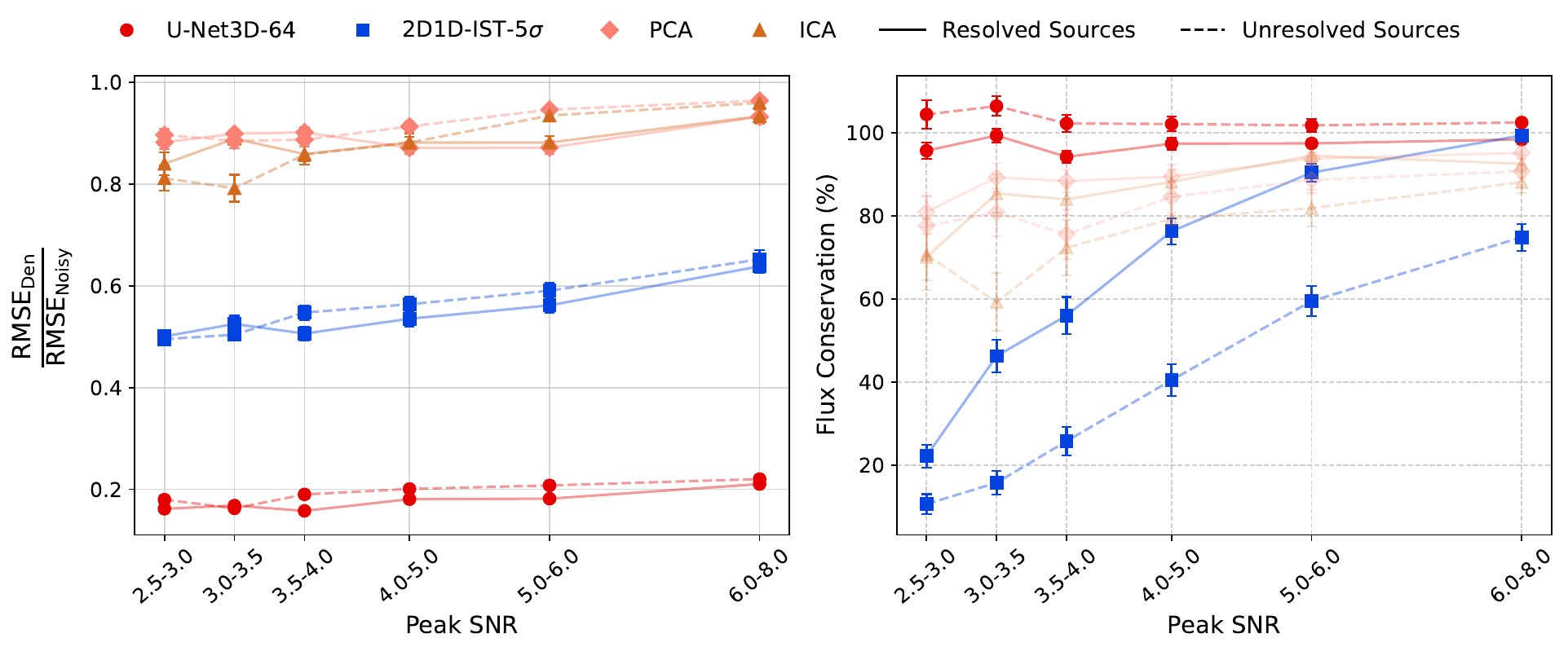}
        \label{fig:flux_rmse_comparison}
    
    \caption{\textbf{(a)} Ratio of the RMSE between denoised cubes and the ground truth to that between the noisy cubes and the ground truth, averaged over 50 cubes and shown as a function of peak SNR and resolution. Solid lines denote sources with resolved emission, while dashed lines indicate unresolved (point-source) emission; \textbf{(b)} Conservation of flux within fixed circular emission apertures, compared across different noise levels and resolutions}
    \label{fig:combined_rmse_figures}
\end{figure*}

\subsection{Comparative Analysis of Toy Cubes}
\label{sec:toy_denoising}

To evaluate the performance of various denoising methods under controlled and reproducible conditions, we first test them on the synthetic toy cubes that we designed to resemble simplified but realistic spectral observations while offering a well-defined ground truth in section \ref{subsection:toy}. This allows for precise and quantitative evaluations of how well each denoising method preserves source flux, spatial morphology, and noise characteristics, particularly in challenging high-noise regimes where emission features are faint and not known \textit{a priori}.

The circular aperture described in section \ref{sec:evaluation} identifies the emission region of the central galaxy in each cube to evaluate the local fidelity of signal recovery, focusing on the trade-offs in RMSE and flux conservation within the apertures containing actual emission, which is important in low SNR regimes where noise dominates the background.

\subsubsection{Data Stratification by Morphology and SNR.} To systematically analyze performance across a range of astrophysically relevant conditions, we divide the full test dataset of 2,000 synthetic cubes into two morphological classes:
\begin{itemize}
    \item \textbf{Unresolved (compact) sources:} Representing galaxies whose emission is confined to a small number of spatial pixels, that is comparable or smaller than the beam of the observation
    \item \textbf{Resolved (spatially extended) sources:} Representing galaxies with emission spread over larger areas, capturing extended morphology after being convolved by the beam.
\end{itemize}

Within each morphological class, we perform binning based on the peak signal-to-noise ratio (SNR) of the sources. The binning strategy is designed to emphasize the low-SNR regime, which is of particular interest for applications to deep field observations:
\begin{itemize}
    \item \textbf{Peak SNR bins:} Finely spaced in the low-SNR regime to capture subtle performance differences, and more coarsely spaced at higher SNRs where performance tends to converge.
\end{itemize}

From each bin (defined by source morphology and peak SNR), we randomly sample 50 cubes to ensure consistent evaluation across the SNR range and morphological diversity. This stratified sampling provides a balanced and fair assessment of denoising performance across different regimes.


The unsupervised and supervised denoising strategies described in sections \ref{sec:unsupervised} and \ref{sec:supervised} are compared:

The performance of each denoising method is evaluated in terms of aperture-based RMSE and flux conservation, both averaged within morphological classes and binned by peak SNR. Our results reveal several nuanced trends:

\subsubsection{Root-Mean-Square Error (RMSE) Ratio within Emission Aperture}\label{sec:rmse_comp}

We first examine the ratio of the RMSE of the denoised cubes to that of the noisy cubes, computed within the fixed circular emission mask defined in section \ref{sec:evaluation}. This metric quantifies the relative improvement in noise suppression, with lower values indicating superior denoising performance.

The PCA and ICA methods yield RMSE ratios close to unity across all noise bins and source morphologies (Figure \ref{fig:combined_rmse_figures}a), indicating minimal noise reduction. This suggests that linear decomposition methods such as PCA and ICA are not well-suited to handling the correlated noise in 3D spectral cubes, which combine spectral and spatial dimensions. Their inability to effectively separate noise from signal under these conditions results in minimal improvement in the RMSE.

In contrast, the 2D-1D wavelet iterative soft thresholding method demonstrates significant noise reduction, achieving RMSE ratios around 0.5-0.6 on average. Notably, the RMSE ratio decreases as the noise level increases (i.e., at lower peak signal-to-noise ratios, SNR), with the best performance observed at peak SNR $\sim 2.5$--$3$. As the noise level decreases (peak SNR rising toward $6-8$), the RMSE ratio increases slightly but remains well below unity. This behavior reflects the wavelet method’s ability to exploit multi-scale representations to attenuate noise while preserving source structures.

The 3D U-Net attains RMSE ratios as low as $\sim 0.18–0.20$ across noise bins, outperforming all other methods. We note that this performance is expected by construction, as the U-Net is trained with a mean-squared error (MSE) loss. Consequently, its low RMSE should be interpreted with this context in mind. While it does demonstrate excellent noise suppression and fidelity with the expected ground truth for previously unseen data, it does not imply that the network is equally optimized for other metrics, such as flux conservation within apertures or recovery of spectral features. In these subsequent evaluations (see sections \ref{sec:flux_comp} and \ref{sec:spectral_features}), the U-Net’s performance reflects its ability to generalize beyond the training objective, providing complementary insight into its effectiveness on aspects of the data not directly targeted by the loss function.

\subsubsection{Flux conservation within Emission Aperture}
\label{sec:flux_comp}

We assess flux conservation by computing the ratio of the integrated flux within the emission mask in the denoised cubes to that in the ground truth. For the U-Net, while the network is trained to directly minimize MSE, and thus achieves low RMSE, this evaluation provides a complementary perspective, allowing us to determine whether the U-Net, as well as the other methods, can preserve physically meaningful quantities such as total emission flux, beyond simply reducing voxel-wise error.

PCA and ICA preserve a large fraction of the flux, close to unity (Figure \ref{fig:combined_rmse_figures}b). However, this apparent flux conservation is primarily due to the lack of significant noise suppression, as indicated by their near-unity RMSE ratios. Thus, the flux is retained along with residual noise, limiting its practical denoising utility.

We note that the wavelet-based methods inherently conserve the total flux since the approximation (or coarse) scale, which captures the non-zero mean component of the signal, is not subject to thresholding. However, aperture-based flux measurements for resolved sources at low peak SNR bins ($\sim2.5$--$3$) can appear significantly suppressed, recovering only $\sim$30--40\% of the true flux. This apparent loss arises because only the wavelet detail coefficients, which are zero-mean by construction, are subjected to thresholding. At high noise levels, significant portions of the faint, spatially extended emission remain undetected in the detail scales due to aggressive thresholding, causing the flux to be retained predominantly in the coarse scale. Consequently, the emission becomes more spatially diffuse, and a substantial fraction of the flux lies outside the designated emission aperture. As the peak SNR increases, thresholding becomes more effective at isolating signal in the detail scales, and the recovered flux within the aperture improves steadily, reaching near-complete recovery ($\sim$100\%) by peak SNR $\sim6$--$8$. For unresolved sources, flux conservation within the aperture remains systematically lower across all SNR bins. This is likely due to the compactness of their spatial structure, which leads to their representation in high-frequency wavelet coefficients that are more susceptible to suppression, resulting in an underestimation of localized flux.

\begin{figure*}[h!]
    \centering
        \includegraphics[width=\textwidth]{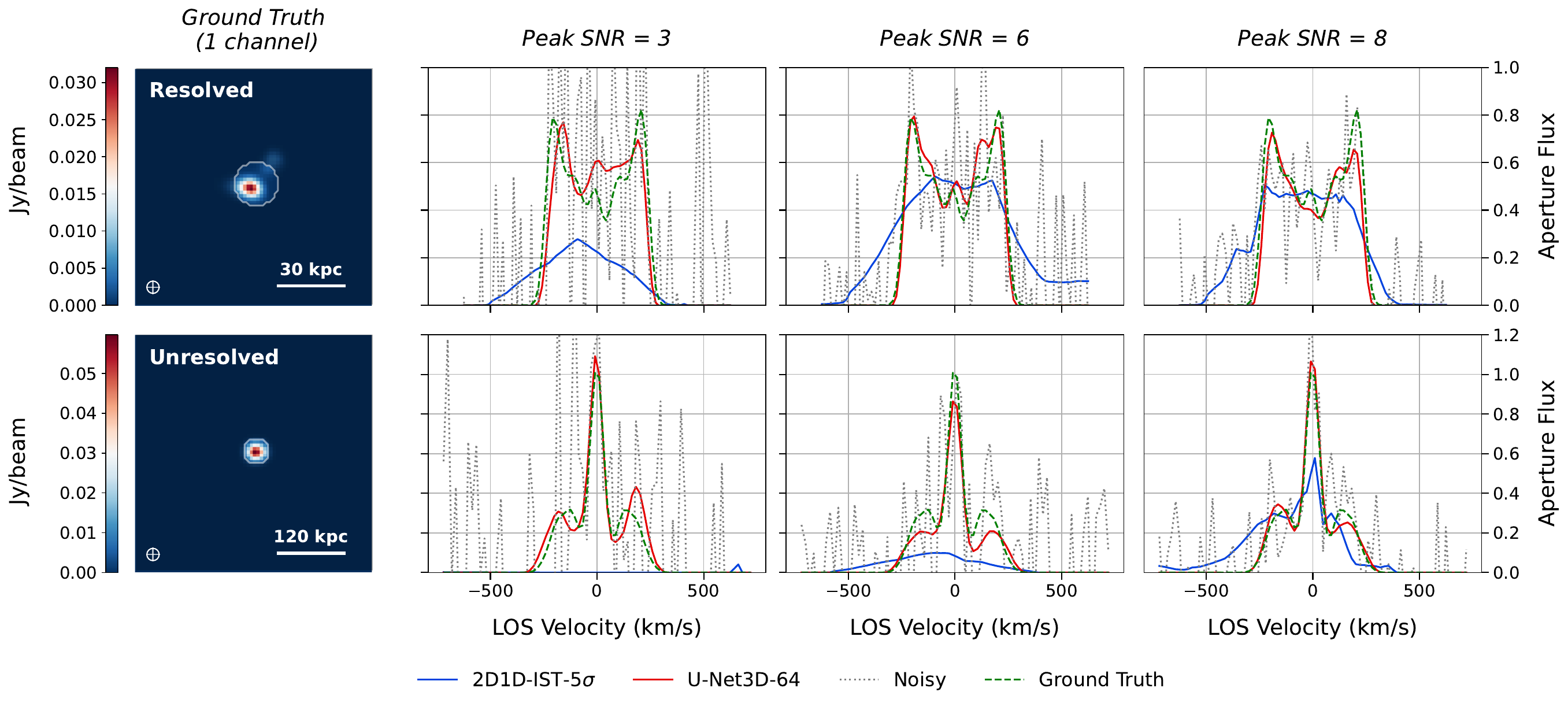}
    
    \caption{Spectral reconstruction for representative \textbf{resolved (top row)} and \textbf{unresolved (bottom row)} sources. The first column shows a ground truth spatial slice with the emission mask overlaid. The next three columns display the ground truth spectrum (green dashed), noisy spectrum (grey dotted), and denoised spectra using 2D-1D-IST (red) and 3D U-Net (maroon) at peak SNRs of 3, 6, and 8, respectively. The resolved source exhibits a double-horned profile, while the unresolved source shows a simpler peak at the systemic velocity.}

    \label{fig:spectra_comparison}
\end{figure*}

The U-Net demonstrates excellent flux conservation for resolved sources, with ratios very close to unity across all noise bins, underscoring its ability to preserve extended source morphology and total flux robustly. However, for unresolved sources, it systematically overestimates the flux by approximately 10\%, particularly at lower peak SNRs (higher noise). This overestimation diminishes as the noise decreases (peak SNR increases), trending toward accurate flux recovery at high SNR.

The systematic flux overestimation for unresolved sources by the U-Net can be attributed to several factors:

\begin{itemize}
    \item \textit{Noise Bias Amplification:} Given the beam-convolved noise, unresolved point sources spread over very few spatial pixels, making it challenging for the network to distinguish between faint residual noise fluctuations and true signal. At high noise levels, the network may interpret noise peaks as emission, thus inflating the integrated flux.
    \item \textit{Spatial Smoothing and Flux Redistribution:} The U-Net’s learned filters can spatially smooth the data, redistributing flux over neighboring pixels. For unresolved sources concentrated in a few pixels, this smoothing may increase the total flux measured within the emission mask, especially if the mask extends beyond the nominal beam size.
    \item \textit{Nonlinear Network Biases:} The network may have learned subtle biases from the training data, particularly regarding unresolved sources embedded in noise, resulting in systematic flux scaling errors more pronounced at low SNRs.
\end{itemize}

The observed trend, where flux conservation improves at higher peak SNRs, arises because the network’s ability to discriminate signal from noise rises as the signal becomes more prominent. This leads to reduced overestimation bias and improved flux recovery in regimes of lower noise.

\subsubsection{Preservation of Spectral Features in Individual Sources}\label{sec:spectral_features}

To further assess the ability of the denoising methods to preserve astrophysically meaningful spectral features, we examine individual examples of a resolved and an unresolved source, as shown in Fig.~\ref{fig:spectra_comparison}. Each row in the figure corresponds to a different resolution regime: the top row presents results for a resolved source with extended emission and well-defined kinematics, while the bottom row shows an unresolved source characterized by compact morphology and a simple spectral profile.

The first column of each row displays a representative spatial slice from the ground truth cube (i.e., prior to the addition of noise), overlaid with the emission mask used for evaluation. The subsequent three columns show the integrated spectra extracted within the emission mask for cubes with added noise at peak SNRs of 3, 6, and 8, respectively. Each spectral panel includes the ground truth spectrum, the noisy spectrum, and the denoised spectra obtained using the 2D-1D Iterative Soft Thresholding (IST) method and the 3D U-Net. We do not include PCA and ICA in this comparison, as earlier analysis (Sections \ref{sec:rmse_comp}, \ref{sec:flux_comp}) showed that these methods do not yield robust denoising performance for our data and noise characteristics.

For the \textbf{resolved source}, the ground truth spectrum exhibits a clear double-horned profile, indicative of rotational kinematic structure. At the highest noise level (peak SNR = 3), the noisy spectrum is heavily corrupted, with significant distortion of spectral shape. The 2D-1D-IST method fails to recover the spectral profile under these conditions, producing a denoised spectrum that is significantly smoothed and poorly aligned with the true structure. As the noise decreases (peak SNR = 6 and 8), the IST method shows some improvement: the overall spectral envelope begins to resemble the ground truth, and the locations of the spectral peaks are partially recovered. However, misalignments persist, and the spectral features remain smoothed. This aligns with the earlier result (Section \ref{sec:flux_comp}) showing that flux conservation for wavelet denoising is poor at low SNRs and improves only gradually, which suggests that spectral detail is also partially lost due to aggressive thresholding at high noise levels. Even at peak SNR $\sim 8$, where flux conservation approaches unity, the subtle misalignment and under-resolution of spectral features persist, pointing to the limitations of the method for recovering detailed kinematic signatures.

The 3D U-Net, on the other hand, performs robustly across all noise levels. Even at peak SNR = 3, it recovers the double-horned structure well, and the fidelity improves with decreasing noise, with correct alignment of peak positions and amplitudes, demonstrating its effectiveness in preserving spectral morphology.

For the \textbf{unresolved source}, the ground truth spectrum is simpler, with a single major peak near the systemic velocity, and minor peaks on either side, as expected for a compact source lacking strong internal kinematic gradients. The noisy spectra again show heavy degradation at low SNRs. The 2D-1D-IST method performs comparably poorly at SNR = 3, failing to reconstruct the spectral shape. However, unlike the resolved case, the denoised spectrum at peak SNR = 8 more closely follows the ground truth shape, albeit with some broadening. This improved performance can be related to the simpler morphology of the unresolved spectrum, likely contributing to the better recovery of the overall trend at high SNRs, though the precision of the reconstruction remains inferior to that of the U-Net.

The U-Net again outperforms the wavelet method, accurately modeling the spectral peak even at high noise levels, and achieving nearly perfect agreement with the ground truth at SNR = 8. Notably, despite the flux overestimation observed for unresolved sources with the U-Net (see section \ref{sec:flux_comp}), the spectral shape is preserved with high fidelity, indicating that the overestimation is likely due to low-level background enhancement rather than distortion of primary emission features.

This analysis illustrates that the U-Net not only improves RMSE and total flux metrics but also excels in preserving detailed spectral structure, which is essential for accurate astrophysical interpretation, especially for kinematically resolved sources. The wavelet-based IST method, while capable of modest flux recovery at moderate SNRs, struggles to reconstruct spectral details, particularly for complex resolved profiles due to its inherent thresholding mechanism.

\subsubsection{Challenges and limitations}

\begin{figure*}[t!]
\includegraphics[width=\textwidth]{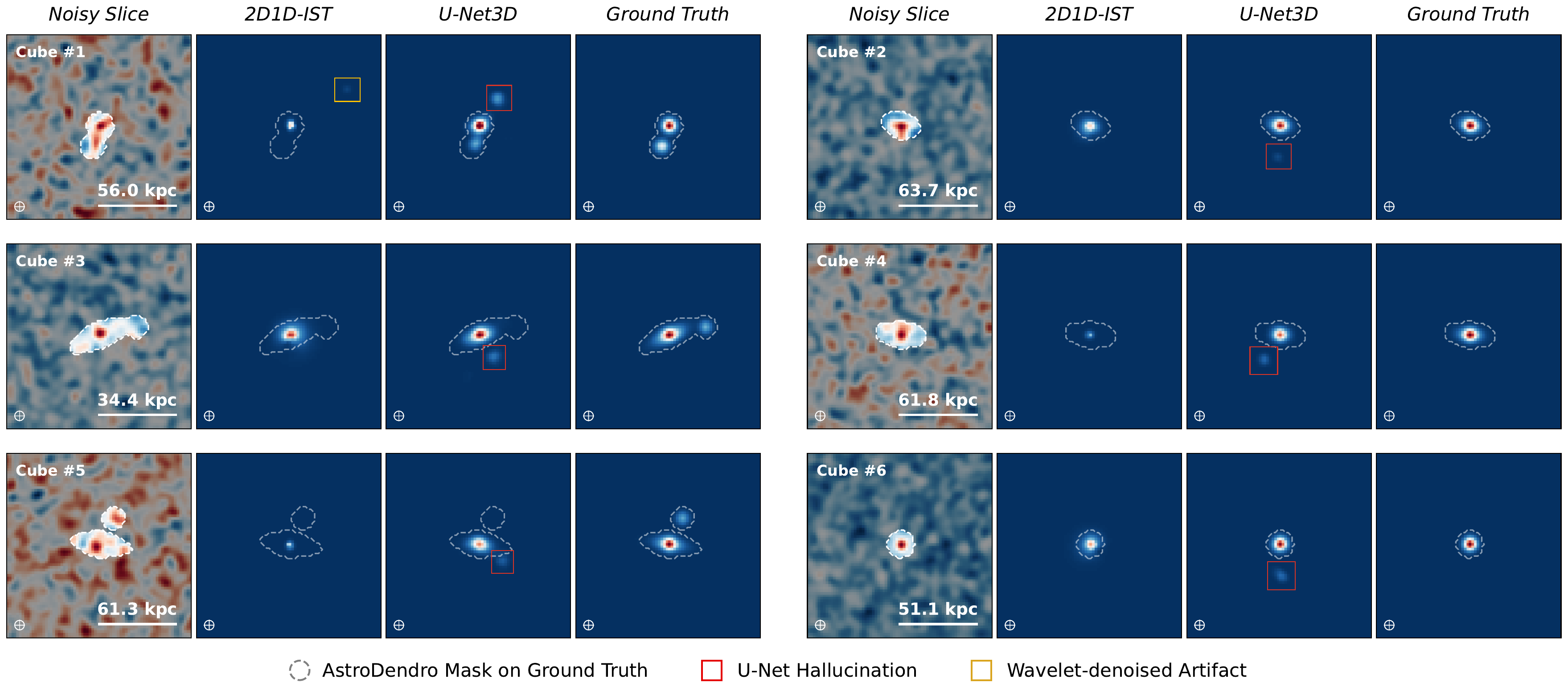}
\caption{Single slices of U-Net denoised cubes for inputs with very high levels of noise. Cyan and black contours represent emission regions identified using AstroDendro on the noise-free ground truth, overlaid on all the subplots to visually depict the precision in the reconstruction of the flux and morphology in the identified regions. The red squares represent hallucinations caused by the neural network while attempting to optimise the RMSE and conserving flux.}
\label{fig:hallucinations}
\end{figure*}

\label{sec:unet_hallucinations}

While the U-Net-based denoising approach performs robustly across the noise range explored in this work and reliably recovers the flux and morphology of the bright central galaxy, it can potentially introduce faint companion-like features absent from the ground truth, i.e., \textit{hallucinations}, indicating a tendency to reproduce morphological patterns learned during training.

In contrast, the wavelet 2D–1D-IST method behaves fundamentally differently, as it does not rely on training data and therefore cannot hallucinate signal in the supervised sense. Its behavior is governed entirely by the noise properties of the data and the threshold applied to the wavelet coefficients. In the presence of strongly correlated noise, clusters of high-magnitude coefficients may survive thresholding and appear as localized statistical anomalies. This effect is threshold dependent: higher thresholds suppress artifacts but risk removing genuine low-flux emission. Throughout this work, we adopt a conservative $5\sigma$ threshold to balance signal retention and anomaly suppression.

Figure~\ref{fig:hallucinations} shows representative single spectral channel map of noisy inputs, noise-free ground truths, and denoised outputs from the wavelet 2D-1D-IST and U-Net methods. Grey contours derived from the ground-truth cubes using \texttt{AstroDendro} are overlaid on all panels to enable direct visual comparison, and U-Net-denoised hallucinations are highlighted with red squares while wavelet-denoised anomalies are highlighted in yellow.

In cubes 2, 4 and 6, that do not have any companion galaxies in the current spectral channel in the ground truth, the U-Net appears to hallucinate spurious structures resembling realistic companions, likely based on training priors of systems with multiple galaxies and the global RMSE obective. In cubes 1, 3 and 6, that show the presence of one companion galaxy in the ground truth, the high level of noise obscures the faint companion, and while the U-Net reconstructs the central source accurately, it fails to recover the full flux distribution, and appears to redistribute missing flux into nearby hallucinated structures. This highlights a fundamental limitation of supervised denoisers: spatial fidelity may be sacrificed for statistical correctness when the signal is extremely weak.

In Figure~\ref{fig:hallucinations}, only cube~1 exhibits a minor threshold-induced anomaly, marked with a yellow square. Examination of the corresponding noisy cube reveals a localized region of high-amplitude correlated noise. All other cubes show no spurious structures, retaining only coefficients associated with the genuine signal.

\begin{figure}[t!]
\includegraphics[width=\columnwidth]{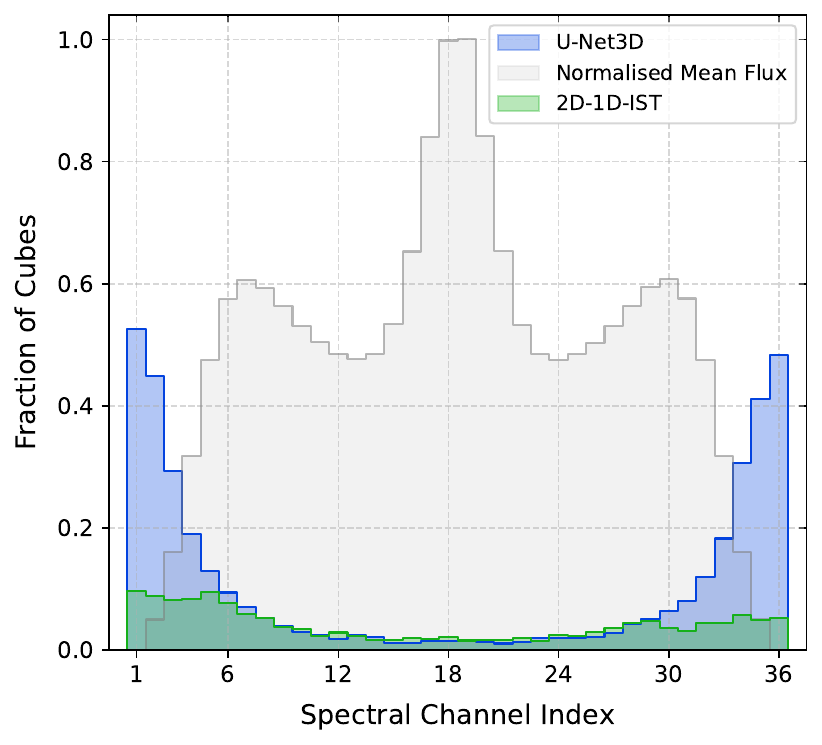}
\caption{Depiction of the number of hallucinating cubes as a function of the spectral channel index for the U-Net3D (blue), wavelet 2D-1D-IST (green), and the normalised average flux of all 2000 cubes (grey).}
\label{fig:hallucination_quant}
\end{figure}

We further quantify these effects across all 2,000 test cubes, we developed a simple, aperture-based methodology to identify coherent hallucinating or anomalous structures (see Appendix~\ref{sec:hallucination_quant} for detailed methodology) per spectral channel, since many channels -- particularly at the velocity extremes -- contain little or no genuine emission and are dominated by noise, making them susceptible to spurious detections. Figure \ref{fig:hallucination_quant} depicts a statistical review of the fraction of cubes where we detect coherent, non-genuine signal. We find that approximately $85.2\%$ of U-Net-denoised cubes exhibit hallucinations in at least one spectral channel, compared to $35.5\%$ of cubes showing at least one statistical anomaly when denoised with the wavelet 2D-1D-IST method. The spectral distribution of these events differs markedly between the two approaches: For the U-Net, $\sim96\%$ of detected hallucinations occur in channels corresponding to velocity extremes, which contain $\leq35\%$ of the mean flux. In contrast, channels with significant genuine emission are largely unaffected, with hallucinations detected in only $\sim4\%$ of cubes. Wavelet-induced anomalies, on the other hand, exhibit a comparatively flatter distribution across velocity channels, indicating a weaker dependence on spectral position. This behavior reflects fundamental differences in how the two methods respond to noise-dominated regions of the data.

\begin{figure*}[t!] \includegraphics[width=\textwidth]{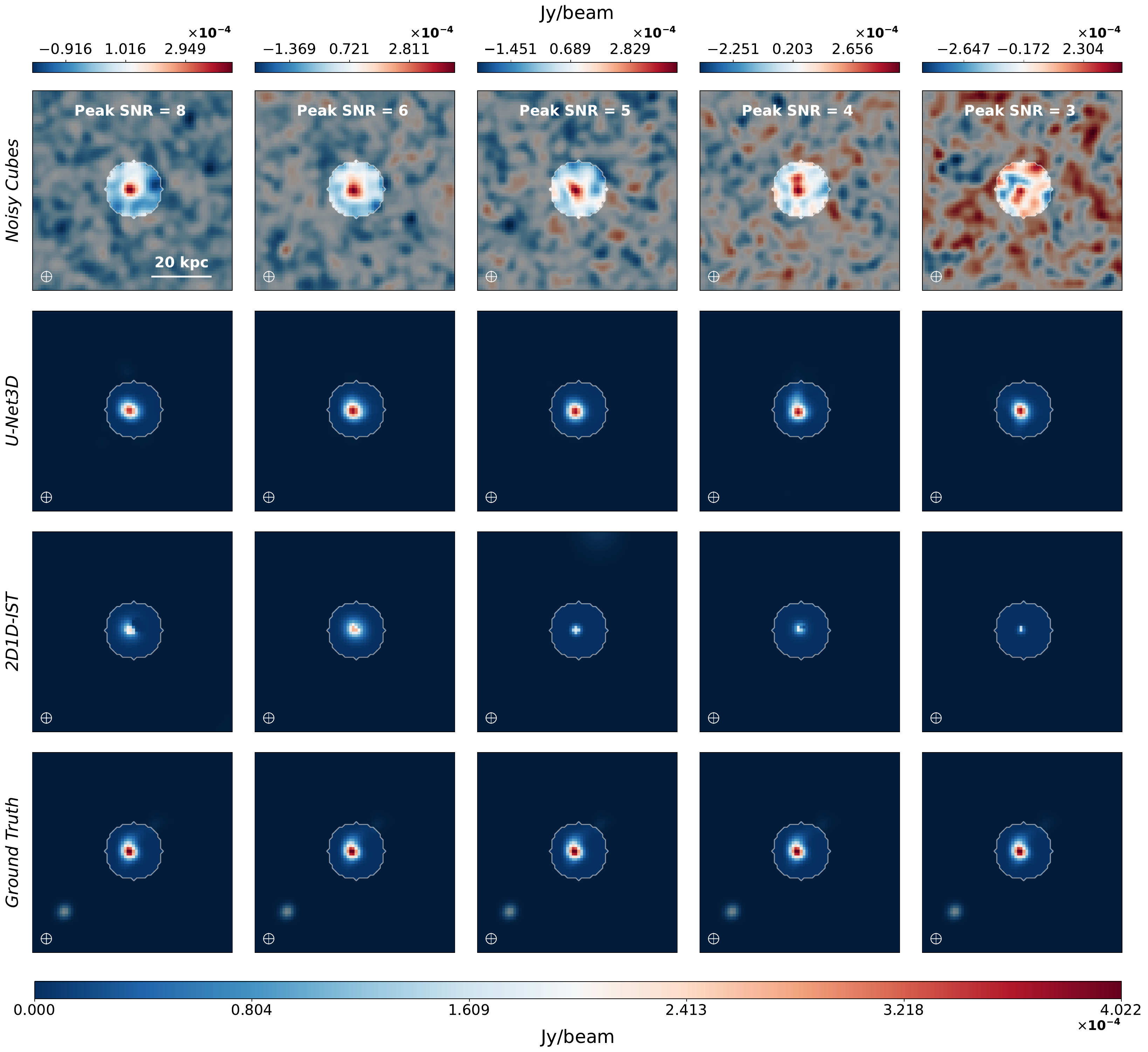} \caption{Different levels of Gaussian noise added into the preprocessed [\ion{C}{ii}] mock IFU cube from the FIRE simulations to test the performance of denoising methods on "realistic" astrophysical data, and the corresponding denoised cubes using Unet3D and iterative 2D-1D wavelet denoising.} \label{fig:sim_cube_denoised} \end{figure*}

To further understand and mitigate U-Net-induced hallucinations, we have performed U-Net stacking by training multiple neural networks with identical initialisation and training data, and visually comparing their outputs on the same examples shown in Figure~\ref{fig:hallucinations} (See Appendix~\ref{A:unet_stacking} and Figure~\ref{A:fig:unet_stacking} for detailed visual analysis). We observe from our subset of examples that while certain hallucinations may seem repetitive across realisations, reflecting model overconfidence in very specific low-SNR regions, hallucinating features present in one U-Net realization can also often be absent in others, or some realisations can have differing hallucinations. This insight suggests that stacking or sampling-based methods - such as averaging predictions from multiple U-Nets, diffusion models, conformal predictions or neural posterior estimators - could be leveraged to statistically quantify model uncertainty robustly and reduce the incidence of spurious signal detection (section \ref{sec:future_work})

\subsection{Application to FIRE Mock IFU Data}
\label{sec:fire_ifu_denoising}

\begin{figure}[t!]
\includegraphics[width=\columnwidth]{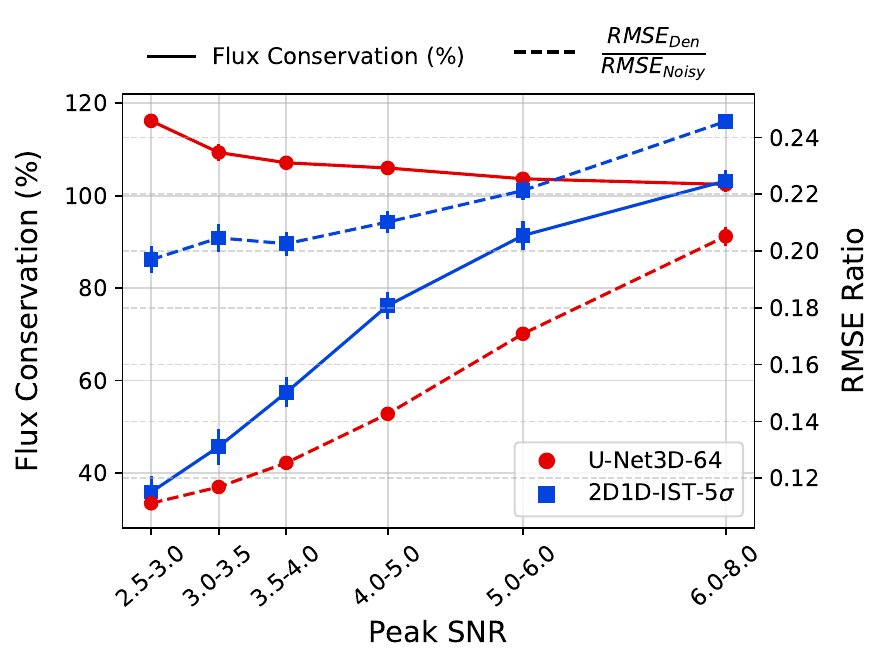}
\caption{Evaluation of denoising performance on the mock FIRE-based IFU cube. Total flux conservation ratio (solid lines, left axis) and normalized RMSE (dashed lines, right axis) are plotted across peak SNR bins for the U-Net3D (maroon) and 2D-1D-IST (red).}
\label{fig:denoised_stats_sim}
\end{figure}

We further test the 3D U-Net and the unsupervised 2D-1D iterative soft thresholding (IST) method on a realistic mock [\ion{C}{II}] IFU cube of a complex, high-redshift multiple-merger system from the FIRE cosmological simulations (Section \ref{sec:mock_ifu}). While this system is astrophysically similar to W2246 in terms of merger-driven dynamics and turbulent kinematics, it is not characterised by the same level of diffuse flux surrounding the host galaxy, instead showing more centrally concentrated emission with some extended structures. The cube incorporates detailed spatial morphologies, irregular spectral profiles, and a known ground truth, providing a stringent test of generalisation for models trained on simpler data. The original high-resolution cube is spatially downsampled via flux-conserving two-dimensional interpolation, and the spectral axis is re-binned to match the shape used for U-Net training, and beam-correlated noise is added with peak SNRs between 2.5 and 8. Figure \ref{fig:sim_cube_denoised} shows noisy, U-Net, and IST outputs across noise levels: visually, the U-Net consistently preserves the central galaxy morphology even at SNR = 3, while IST retains flux at high SNR but is increasingly unable to recover faint structures within the aperture at lower SNR.

For quantitative evaluation (Fig. \ref{fig:denoised_stats_sim}), we generate 200 noise realisations per SNR bin and compute the RMSE within the fixed emission mask, normalised by the noisy cube RMSE, along with the denoised-to-true total flux ratio; negligible error bars indicate tightly constrained means. Across all SNRs, the U-Net achieves the lowest RMSE, with values increasing toward high SNRs, where there is less room for improvement. Meanwhile, IST follows the same trend but with consistently higher RMSE, indicating weaker noise suppression.

In terms of flux recovery, the U-Net is robust at high SNR but tends to progressively overestimate flux at low SNR, likely due to beam-correlated noise being interpreted as signal and training-set biases favoring slightly higher flux in noisy compact sources. IST conserves flux well at high SNR but increasingly underestimates it at low SNR as its 5$\sigma$ thresholding removes genuine faint emission where signal and noise overlap. These results show that a U-Net trained entirely on idealized toy cubes can generalise well to realistic IFU data, preserving morphology and spectra while recovering flux reliably aside from known low-SNR caveats.

This bridges the gap between synthetic and observational applications, demonstrating that large volumes of inexpensive toy simulations can enable fast, high-quality denoising of expensive, noise-limited datasets from facilities such as ALMA, JWST, and VLT/MUSE, while the IST method provides a physically interpretable benchmark whose limitations for faint emission at low SNR remain evident.

\begin{figure*}[t!]
    \includegraphics[width=\textwidth]{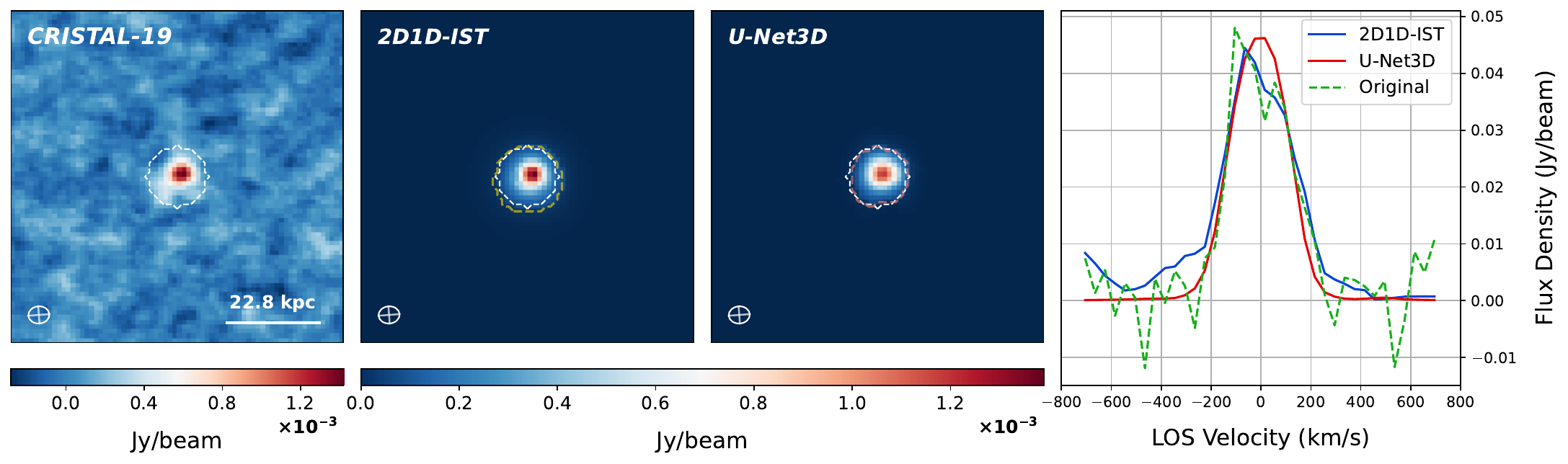}
    \includegraphics[width=1.005\textwidth]{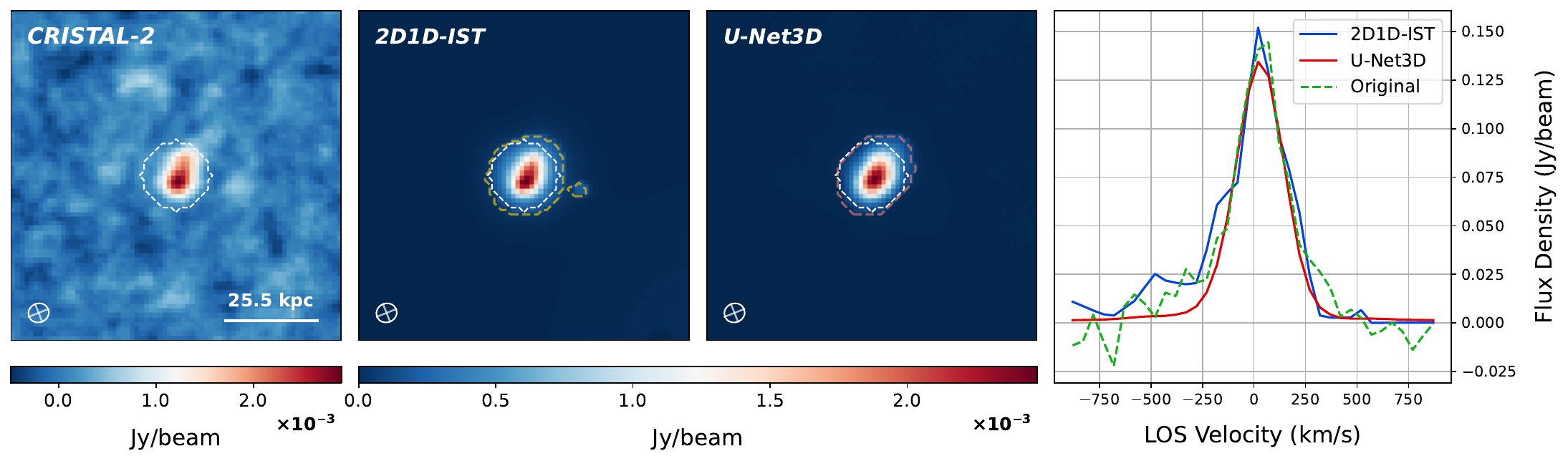}
    \caption{Denoising performance of CRISTAL-19 and CRISTAL-2 IFUs (top and bottom rows respectively) within the emission aperture, by 2D-1D-IST and U-Net, and the associated noisy and denoised spectra.}
    \label{fig:cristal_test}
\end{figure*}

\subsection{Application to \textit{CRISTAL} Spectral Cubes}
\label{sec:cristal_result}

We applied both the iterative 2D-1D wavelet soft-thresholding method and the supervised 3D U-Net to [C\,\textsc{ii}] spectral cubes from the CRISTAL survey, focusing on CRISTAL-19 and CRISTAL\,2 (Section~\ref{sec:cristal}). CRISTAL-19 is spatially unresolved by the synthesized beam, while CRISTAL-2 is resolved. Both targets have peak SNRs of $\sim15$, comfortably within the lower-noise regime where the 2D--1D wavelet approach has previously been shown (Section~\ref{sec:toy_denoising}) to recover most of the flux for higher peak-SNR sources ($\gtrsim6$--8) without any pre-processing or supervised training.  

The averaged CRISTAL-19 beam has a major axis FWHM of 0.71", corresponding to $\sim4.45$ kpc in the total observed field of view of 180.52 kpc on both axes, and 7.03 px in the observed pixel resolution; the minor axis has a FWHM of 0.58" $\sim3.64\text{ kpc }\sim 6.34\text{ px}$; and the beam position angle is $93.72^\circ$, defined as the orientation of the major axis measured east of celestial north. For the U-Net application, to ensure architectural and physical compatibility, we preprocess the data cube such that the FWHM of the beam minor axis is consistent with the FWHM of the circular beam used for training in pixel units (3.75 px) while downscaling the pixel grid to $72\times72\:\rm px^2$: The original cube was first symmetrically spatially cropped around the source galaxy to a grid of $122\times122\rm\:px^2$ spanning a field of view $\sim 69.97$ kpc on both axes, and downsampled by a factor $\sim1.69$ using bilinear interpolation (final beam minor axis FWHM $\sim 3.74$ px). As we have trained our U-Net with cubes with variable spectral/line-of-sight velocity resolutions within 36 channels, the CRISTAL-19 spectral axis, originally comprising 200 channels with a velocity resolution of 10 km/s was symmetrically cropped to the nearest multiple of 36 (180 channels) and subsequently averaged along the spectral dimension.

Similarly, CRISTAL-2 with average beam specifications of 0.57" $\times$ 0.51" $\sim$ 3.56 kpc $\times$ 3.18 kpc $\sim$ 7.03 px $\times$ 6.28 px in a field of view of (202 kpc)$^2$ with a beam position angle of $109^\circ$, is symmetrically cropped to a grid of (121 px)$^2$ and downsampled by a factor of 1.68, and spectral rebinning is performed.

For a noisy cube $\bm{Y}$ and corresponding denoised cube $\bm{X}^{\rm den}$, the flux conservation ($S(\bm{X}^{\rm den}_{\in \mathcal{A}})/S(\bm{Y}_{\in \mathcal{A}})$) percentage was quantified within a fixed circular aperture defined on the noisy cube ($\mathcal{A}$). For CRISTAL-19, the U-Net conserved $\sim94\%$ and the wavelet method preserved $\sim100\%$ of the original flux and recovering $\sim16$\% additional flux. For CRISTAL-2, the U-Net conserved $91.6\%$ and the wavelet method conserved $\sim100 + 15\%$.

Although the initial noise in the cubes is Gaussian, both the U-Net and the wavelet-based denoising methods are inherently non-linear. As a result, the denoised cube does not retain a Gaussian distribution of noise levels. In particular, the wavelet method enforces sparsity in the signal representation, which further introduces strong non-Gaussianity. To mitigate these effects and obtain a robust quantification of noise suppression, we compute the residual cube:
\begin{equation}
\bm{R} = \bm{Y} - \bm{X}^{\rm den},
\end{equation}

We then quantify the percentage of noise reduction by computing a ratio between the ratio of the MAD-estimated noise levels (see equation \ref{eq:mad}) per beam of the residual cube to that of the original noisy cube, computed outside the circular aperture ($\notin \mathcal{A}$):
\begin{equation} \label{eq:noise_reduction}
\mathrm{Noise~Reduction} = \frac{\sigma_{\rm res}}{\sigma_{\rm noisy}} = \frac{\sigma_\mathrm{MAD}(\bm{R}_{\notin \mathcal{A}})}{\sigma_\mathrm{MAD}(\bm{Y}_{\notin \mathcal{A}})} \times 100 \%,
\end{equation}
In this formulation, a value of 100\% indicates perfect noise removal, whereas 0\% corresponds to no noise being removed.

Both the U-Net and 2D-1D wavelet methods achieved strong noise suppression in the cubes, with the residual noise reduced by $\gtrsim 99\%$ relative to the original noise.

We can approximately calculate the improvement in SNR: The MAD noise standard deviation of the denoised cube can be approximately computed as:

\begin{equation}
    \sigma_{\rm den} = \sigma_{MAD}(\bm{X}^{\rm den}_{\notin \mathcal{A}}) \approx \sqrt{\sigma^2_{MAD}(\bm{Y}_{\notin \mathcal{A}}) - \sigma^2_{MAD}(\bm{R}_{\notin \mathcal{A}})}
\end{equation}

as the variance scales linearly. Consequently, the improvement fraction of SNR can be described as:

\begin{equation}\label{eq:snr}
\text{SNR Improvement} 
\approx \dfrac{S_{\rm den}/\sigma_{\rm den}}{S_{\rm noisy}/\sigma_{\rm noisy}}.
\end{equation}

Where $S_{\rm den}$ and $S_{\rm noisy}$ are the total flux observed within the emission aperture in the denoised and noisy cubes, respectively. Therefore, for CRISTAl-19, the U-Net observes an approximate SNR improvement by a factor of $\sim 11.7$, and the wavelet 2D-1D improvement factor is $\sim 13.5$. For CRISTAL-2, the approximate SNR improvement factors for U-Net and wavelet denoising strategies are $\sim 6.9$ and $\sim 11.1$, respectively.

These results are consistent with our synthetic toy-cube experiments, where the U-Net typically outperformed the 2D-1D IST method in RMSE and flux conservation across a range of noise levels, but the wavelet method retained most of the flux for high-SNR regimes (peak SNR $\ge 6-8)$. The CRISTAL cubes satisfy this high-SNR condition, and despite their far more complex morphologies and spectra compared to the toy data, they confirm the robustness of the 2D-1D wavelet method as a purely unsupervised method for IFU observations in this regime. It also provides a fascinating demonstration of the robustness of a U-Net trained on a purely synthetic dataset with real observational data.

In the following subsection, we contrast these results with those of W2246-0526, a low-SNR, dynamically disturbed system with diffuse emission, which presents a more challenging denoising scenario and tests the robustness of both methods under markedly different physical conditions.

\subsection{Testing on Complementary Sample: \textit{W2246–0526}}
When testing on the W2246 [C\,\textsc{ii}] cube at high SNR ($>$ 50 peak SNR), our goal was to examine how the denoising methods behave under markedly different physical conditions than those present in the datasets used for training, validation, and CRISTAL. W2246 is characterised by extensive diffuse emission surrounding the central host galaxy (\ref{sec:w2246_data}), in sharp contrast to the compact or moderately extended emission structures in our synthetic and CRISTAL datasets. 

Similar to section \ref{sec:cristal}, we preprocessed the spectral cube with beam specifications of 0.45" $\times$ 0.36" $\sim$ 3.0 kpc $\times$ 2.4 kpc $\sim$ 8.9 px $\times$ 7.2 px in an initial pixel grid of $\rm(600\;px)^2$ spanning a field of view of $\rm(200.4\;kpc)^2$, by spatially cropping the data cube symmetrically around the central galaxy to a grid size of $\rm(138\;px)^2$ spanning a field of view of $\rm(46.1\;kpc)^2$, and using bilinear interpolation to downsample the spatial dimensions by a factor of 1.92 for the FWHM of the beam minor axis (3.74 px) to correspond with the FWHM of the circular training beam (3.75 px). The spectral axis is rebinned accordingly to maintain architectural compatibility.

\begin{figure*}[t!]
    \includegraphics[width=\textwidth]{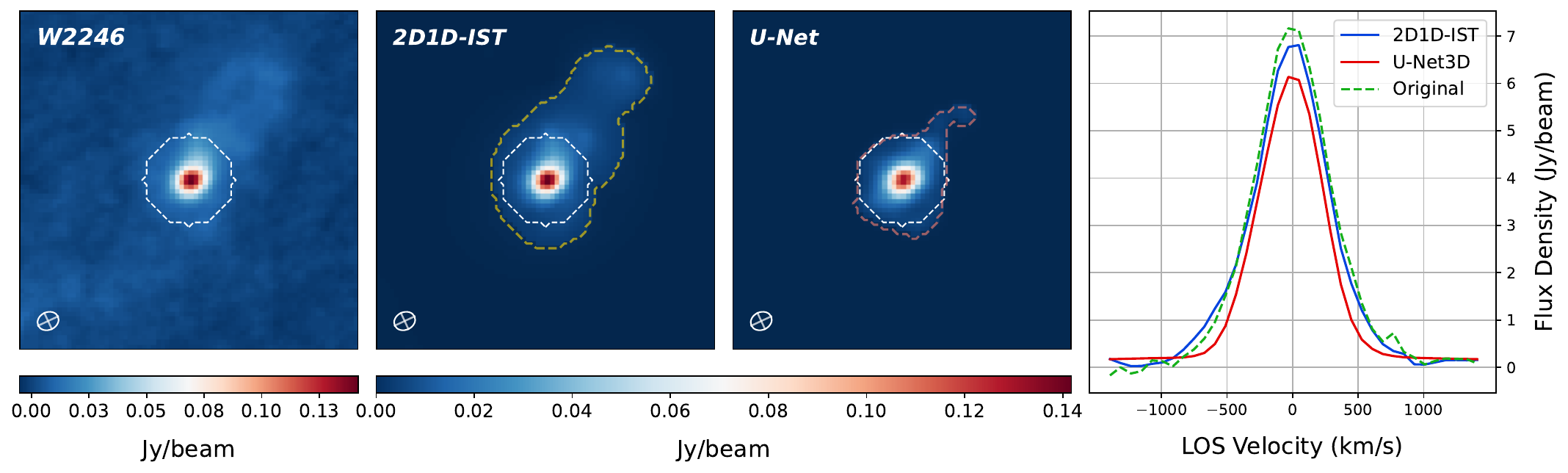}
    \caption{Denoising performance of W2246 within the emission aperture, by 2D-1D-IST and U-Net, and the associated noisy and denoised spectra.}
    \label{fig:w2246_test}
\end{figure*}

The 2D-1D IST wavelet method, which has proven highly effective at high SNR, recovers $\sim95.8$\% of the flux both in and around the host galaxy within the fixed circular aperture, including the extended diffuse emission, as identified by the AstroDendro mask in figure~\ref{fig:w2246_test}. This is plausibly aided by the fact that the coarse wavelet scale is not thresholded, allowing large-scale, low-frequency flux to be preserved. In contrast, the U-Net, trained exclusively on toy rotating galaxy cubes without diffuse emission, has only encountered extended emission in the form of large effective radius central galaxies. Interestingly, despite having significant physical differences between the training set and W2246, the U-Net is able to conserve $\sim$80\% of the flux within the circular aperture, however it is unable to recover diffuse emission and morphology around the central galaxy, as depicted by the AstroDendro mask in figure~\ref{fig:w2246_test}. The corresponding spectra further reflect this: while the wavelet method closely follows the original spectrum, the U-Net depicts an uunderestimation in flux despite preserving the overall spectral shape.

We use a similar methodology as in section \ref{sec:cristal_result} (equation \ref{eq:noise_reduction}) to quantify the percentage of noise reduction. To account for the extended diffuse emission, we manually identified an emission-free spatial area in the cube containing only the background noise and calculated the MAD-estimated standard deviation of the residual and the input cube in that region. Both methods can suppress background noise significantly, with the U-Net achieving $\sim$99\% reduction of noise and the wavelet method achieving $\sim$95\%.

We also quantify the approximate SNR improvement (\ref{eq:snr}) for the two methods only within the central emission aperture - the U-Net observes an increase in central galaxy SNR by a factor of $\sim5.3$, while the wavelet 2D-1D-IST method observes a $\sim3$ times increase.

From a physical and methodological standpoint, this outcome is unsurprising. The U-Net's training regime, although diverse in galaxy size, inclination, and noise conditions, did not encompass morphologies with significant diffuse emission or irregular, merger-driven kinematics. Given that supervised deep learning models strongly depend on their training priors, the absence of such structures in the training set naturally limits the network's ability to reconstruct them at inference time. Incorporating examples with realistic diffuse flux into the synthetic training set or fine-tuning the model with a small subset of realistic data could mitigate this limitation, improving the network's generalisability to complex sources such as W2246. Nonetheless, the fact that the model can recover up to $\sim 80\,\%$ of the aperture flux, demonstrates a significant degree of robustness in its learned representation of central galaxy emission. In practice, this suggests that while wavelet-based methods may be more reliable for total flux recovery in sources dominated by extended low-surface-brightness features, a suitably augmented U-Net could close this performance gap while retaining its strengths in denoising compact and moderately extended structures.

\section{Conclusions and Future Directions}
\label{sec:conclusions}

In this work, we present a comprehensive, multi-tiered evaluation of denoising strategies for spectral cubes of high-redshift sources. Our framework spans a wide range of data realism, from fully controlled synthetic toy models to state-of-the-art cosmological simulations to real ALMA observations. We benchmark the performance of four denoising methods - Principal Component Analysis (PCA), Independent Component Analysis (ICA), iterative 2D-1D wavelet soft thresholding (IST), and a supervised 3D U-Net - on preserving astrophysical signal across varying levels of noise and spatial resolution.

A central contribution of this study is the development of a highly configurable and physically motivated synthetic dataset. The toy spectral cubes are constructed from rotating disk galaxy models with adjustable kinematic profiles, Sérsic intensity distributions, inclination angles, companion systems, beam convolution, and correlated Gaussian noise. Every relevant parameter, including beam size, spatial resolution, and noise level, is explicitly controlled. This flexibility allows us to explore a wide variety of scenarios and generate statistically robust ensembles of test cases\footnote{Open-source software developed: \texttt{GalCubeCraft}\\ \href{https://github.com/arnablahiry/GalCubeCraft}{Public GitHub Repository}, \href{https://arnablahiry.github.io/software/GalCubeCraft}{API Documentation}}. Notably, these cubes serve as ideal training sets for supervised learning, with precise ground truth knowledge and coverage of morphologies and SNRs that span real observational conditions. 

We use a single mock IFU from the zoom-in FIRE cosmological simulations, focusing on a high-redshift system analogous to W2246; however, lacking the same amount of diffuse emission. This serves as a bridge between our toy models and real observations, as this provides a realistically modeled ground truth to validate denoising performance.

We utilize ALMA [C\textsc{ii}] IFUs from the CRISTAL survey and focus on 2 samples, CRISTAL-19 and CRISTAL-2, which are unresolved and resolved by the beam, respectively. The CRISTAL samples serve as the primary observational reference, as the chosen sources are compact rotating galaxies with no diffuse emission. We also consider the hyperluminous quasar W2246-0526 as a complementary observational reference to test the robustness of our methods on a test source that has a markedly different morphology and physics, as evident by the significant diffuse flux surrounding the host galaxy.

Our findings reveal several key insights:

\begin{itemize}
    \item \textbf{Classical statistical methods (PCA, ICA)} demonstrate limited utility in the context of 3D interferometric data. Their linear assumptions and reliance on second-order statistics make them ill-suited for separating signal from spatially correlated noise, particularly in low SNR regimes. This is evident in the toy cube experiment due to the high ratio of the RMSE of the denoised cubes to the RMSE of the noisy cubes with respect to the ground truths.
    
    \item \textbf{The 2D-1D wavelet-based iterative soft thresholding} offers a physically interpretable and unsupervised approach that excels in preserving sharp, compact features and total flux within the emission aperture in medium-to-high SNR regimes. The multi-scale decomposition allows effective noise suppression while respecting the anisotropic structure of spectral cubes. However, at low SNR, the method tends to disregard faint, diffuse emission, particularly in unresolved sources, due to conservative thresholding, as the flux is not retained as signal and instead is retained by the coarse scale, which is not thresholded. This diffuse signal suppression effect was similarly observed in the analysis of the FIRE mock IFU. Tests with both CRISTAL samples proved the robustness of this unsupervised method by conserving >95\% of the total flux within the aperture while aggressively denoising the cube, increasing the approximate SNR by factors $\sim 6.9-7$. This method is also able to recover $\sim100\%$ of the expected aperture flux, and additional flux recovery; including the diffuse emission from W2246, while increasing the already-high SNR by a factor of $\sim 3$.
    
    \item \textbf{The 3D U-Net, trained exclusively on our synthetic toy dataset}, exhibits distinct performance characteristics depending on the regime. In \textbf{low-SNR simulated environments}, it delivers the highest overall fidelity and lowest RMSE, successfully recovering structures where unsupervised methods struggle. However, in \textbf{moderately-high-SNR observational regimes} with morphologies similar to the training data (e.g., CRISTAL), its performance is comparable to the 2D-1D wavelet method, with both recovering $>90\%$ of the total flux. For \textbf{morphologically complex systems} that diverge significantly from the training distribution, such as W2246-0526, the U-Net slightly underperforms the wavelet-based method - recovering $\sim80\%$ of the flux of the central galaxy and immediate surroundings as constrained by the circular aperture, however in contrast to the 2D-1D-IST method, it is unable to recover any external diffuse emission. This underscores that while the U-Net generalizes impressively to realistic/real data, its reliability is bounded by the morphological priors learned during training, whereas the wavelet method remains robust to exotic structures.

    \item \textbf{Hallucinated features} are a notable limitation of supervised deep learning denoisers like the 3D U-Net in low-SNR regimes, where spurious satellite-like structures occasionally arise in previously unseen test data, reflecting learned priors rather than data fidelity. This flux redistribution highlights a trade-off between minimizing loss and preserving spatial accuracy, emphasizing the need for interpretability and uncertainty quantification.

\end{itemize}

The iterative reweighted 2D-1D wavelet soft-thresholding method represents a robust and physically interpretable baseline for denoising spectral cubes with peak SNRs $\ge$ 6–8. This approach has demonstrated consistent performance across our synthetic toy models, FIRE mock IFUs, and CRISTAL observations, yielding accurate flux recovery without introducing anomalous or spurious features when a conservative threshold is applied. Its transparency derives from the use of analytically defined wavelet functions, providing a clear physical basis for the decomposition. However, it can have a subtle misalignment in recovered spectra and loss of kinematic signature even at peak SNR $\sim6-8$, which is mitigated for higher peak SNRs like the CRISTAL samples.

The overall effectiveness and fidelity expressed by the U-Net in the denoising of synthetic as well as observed spectral cubes of similar underlying physics demonstrates that a network trained purely on synthetic data can generalize to denoise complex, realistic spectral cubes, preserving flux and suppressing noise, while underscoring the transformative potential of supervised learning and the value of well-constructed simulations: their controllability, scalability, and inherent access to ground truth make them indispensable for training robust denoisers in the absence of extensive clean observational datasets.

This naturally leads to a broader implication: our toy dataset can serve as the first stage in a transfer learning framework, where models are initially trained on large, generalized synthetic data to capture broad spatial-spectral correlations, and subsequently fine-tuned on small samples of task-specific real/realistic astrophysical data that are typically limited due to computational expense or scarcity in construction or observation at a large scale to incorporate realistic complexity while mitigating the limitations.\\

\subsection{Future work}\label{sec:future_work}
\begin{enumerate}
    \item Incorporating physically motivated priors into training data using cosmological simulations with complex feedback and radiative transfer.
    \item Introducing uncertainty-aware objectives and probabilistic denoisers that quantify confidence in the reconstructed signal; integrate hallucination detection algorithms leveraging methods such as \textit{Quantile Regression} \citep{Xu_2023} and \textit{Conformal Predictions} \citep{MAL-101}, which provide statistically rigorous uncertainty bounds to identify spurious features and improve the reliability of denoised outputs. This has been successfully applied in the context of weak lensing mass mapping by \citet{leterme2025}.
    \item Extending the U-Net-based diffusion model plug-and-play frameworks \citep[e.g.,][]{feng2024, dia2025} used previously for single-channel images to three-dimensional data as a potential solution to the challenges of uncertainty quantification and hallucination in spectral cube denoising tasks.
    \item Developing a hybrid denoising framework that combines the interpretability of sparse models (e.g., wavelets) with the representational power of deep neural networks - \textit{Learnlets} \citep{Ramzi2023-ih}: have architectures that learn multiscale wavelet-like filters directly from data, retaining the sparsity and localization properties of wavelets while benefiting from data-driven learning for improved denoising performance. Previously used by \citet{utsav} for galaxy image deconvolution.
    \item Exploring regularization techniques and loss functions that penalize flux bias and hallucination in low-SNR regimes.
\end{enumerate}

With the advent of deeper ALMA surveys, JWST IFUs, VLT/MUSE, and future surveys, noise suppression will be a fundamental challenge in extracting faint spectral signals from the early Universe. The methodology proposed in this work - combining toy simulations, mock IFUs, and real observations within a unified evaluation framework—offers a pathway toward generalizable, trustworthy, and high-performing denoising pipelines for modern astrophysics.

\begin{acknowledgements}
This work is supported by the TITAN ERA Chair project (contract no. 101086741) within the Horizon Europe Framework Program of the European Commission, and the Agence Nationale de la Recherche (ANR-22-CE31-0014-01 TOSCA).

DAA acknowledges support from NSF CAREER award AST-2442788, STScI JWST grants AR-04357.001-A and AR-05366.005-A, an Alfred P. Sloan Research Fellowship, and Cottrell Scholar Award CS-CSA-2023-028 by the Research Corporation for Science Advancement.

The authors thank the anonymous reviewer for helpful comments and suggestions.

\end{acknowledgements}

\bibliographystyle{aa}
\bibliography{references}

\onecolumn
\appendix

\section{Schematic diagram of the U-Net Architecture}

\begin{figure}[h!]
    \centering
    \includegraphics[width=0.9\textwidth]{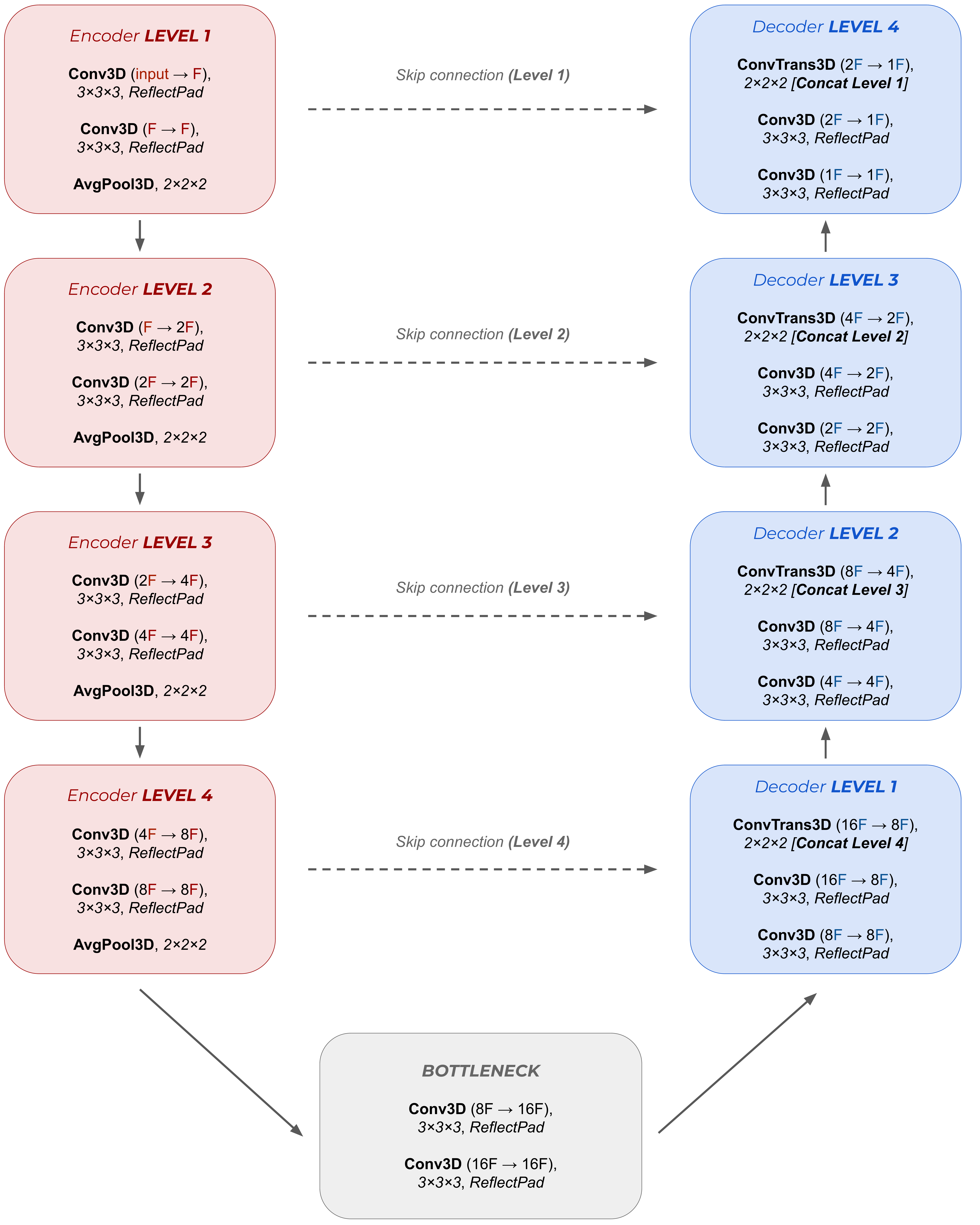}
    \caption{Simple schematic of the three-dimensional U-Net architecture used in this study. \textbf{Four encoder blocks} - each consisting of two 3D convolutions with kernel size 3 and \textbf{reflective padding to avoid edge artifacts}, followed by LeakyReLU activations—perform feature extraction. \textbf{Downsampling} is implemented using \textbf{average pooling}, which preserves the mean surface brightness of the input cubes (expressed in $\rm Jy\;beam^{-1}$) during resolution reduction. \textbf{Four decoder blocks} each apply a 3D transposed convolution with kernel size 2, concatenate the corresponding encoder-block features through \textbf{skip connections}, and apply two further 3D convolutions with kernel size 3.}
    \label{fig:visualise_unet}
\end{figure}

\twocolumn

\section{Hallucination Analysis}\label{sec:hallucination_quant}
\subsection{Quantification}

We analyse and quantify coherent hallucinations in U-Net denoised spectral cubes ($\bm{X_{NN}}(\mathbf{x}, v)$, where $\bm{x}$ is the position vector and $v$ is the velocity channel) and statistical anomalies in wavelet 2D-1D-IST-denoised spectral cubes ($\bm{X_{IST}}(\mathbf{x}, v)$) for $2\times10^3$ cubes of the test dataset, 

To identify astrophysically meaningful emission, we use \texttt{Astrodendro} to express the data as hierarchical trees of connected structures based on flux density. This approach allows us to systematically capture coherent emission regions across both spatial and spectral dimensions. For the ground-truth data cube, a dendrogram is computed to extract the most significant and extended structures. All main branches of the dendrogram are combined to form an emission mask/aperture, which defines the expected spatial-spectral support of true emission ($M_{\rm true}$), which serves as a baseline for evaluation. The same methodology is used to produce emission apertures for the denoised cubes for both methods ($M_{\rm den}$). These apertures identify coherent structures present in the denoised outputs, which can then be compared to the reference aperture to detect spurious structures or statistical anomalies introduced by the denoising process.

Hallucinated (or anomalous) emission is defined as coherent structures identified in the denoised cube but absent from the clean emission mask.

For a denoised cube, the hallucination mask at velocity slice $v$ is
\begin{equation}
\label{eq:hal_structures}
    M^{(v)}_{\rm out} = M^{(v)}_{\rm den} \cap \neg M^{(v)}_{\rm true},
\end{equation}

We define an aperture flux-based hallucination index
\[
R^{(v)} =
\frac{
\sum_{\mathbf{x} \in M^{(v)}_{\rm out}}
{X^{(v)}_{\rm den}}
}{
\sum_{\mathbf{x} \in M^{(v)}_{\rm den}}
{X^{(v)}_{\rm den}}
},
\]

This metric satisfies:
\begin{itemize}
    \item $R \rightarrow 0$: negligible hallucination as there is no coherent flux detected outside the reference emission aperture.
    \item $R = 0.5$: 50\% of the recovered coherent flux is hallucinated.
    \item $R \rightarrow 1$: dominant hallucination as the coherent flux detected outside the ground truth reference aperture is at the same magnitude as the total flux detected in the denoised output.
\end{itemize}

We adopt a conservative threshold
\[
R_{\rm thresh} = 0.1,
\]
and classify a slice as hallucinated or anomalous when $R > R_{\rm thresh}$.

We quantify the presence of spurious emission structures, or hallucinations, in each velocity slice. For a given slice $v$, we compute the fraction of cubes exhibiting hallucinations or statistical anomalies as

\[
f(v) = \frac{N^{(v)}_{\rm hall/anom}}{N_{\rm cubes}},
\]

where $N^{(v)}_{\rm hall/anom}$ is the number of cubes for which coherent structures are detected outside the reference emission mask in the velocity channel $v$, and $N_{\rm cubes}$ is the total number of cubes in the test dataset.  

\begin{figure}[t!]
\includegraphics[width=\columnwidth]{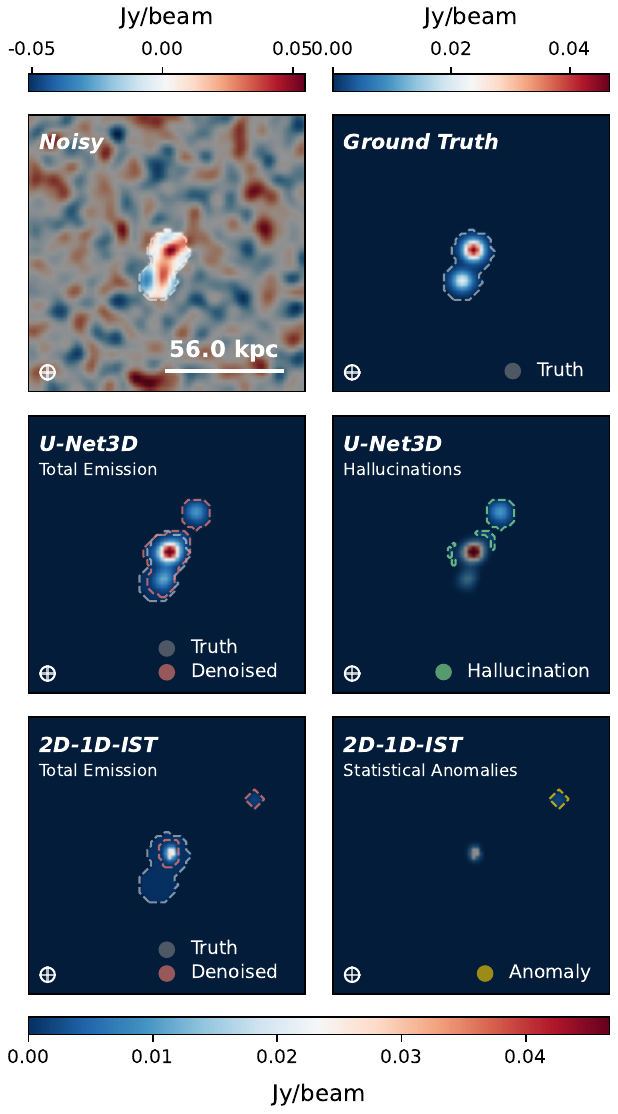}
\caption{Visual depiction of identification of coherent U-Net-induced hallucinating structures and wavelet-induced statistical anomalies in one slice of cube 1 of the examples used in section \ref{sec:unet_hallucinations} with the methodology of equation \ref{eq:hal_structures}}
\label{A:fig:hallucination_masks}
\end{figure}

This slice-wise metric provides a simple yet reasonably robust, physically motivated, and model-agnostic framework for tracking the prevalence of spurious coherent structures as a function of velocity channel and for comparing the behaviour of different denoising methods. The analysis is performed independently for each spectral slice, motivated by the fact that many channels—particularly toward the velocity extremes—contain little or no genuine emission and are therefore dominated by noise, making them especially susceptible to false detections. By assessing the propensity for hallucinated or statistically anomalous structures across the spectral dimension, and by grounding the statistic in dendrogram-defined emission weighted by recovered flux rather than binary detection, this approach enables a nuanced characterization of hallucinations. For more accurate and uncertainty-aware quantification78uyttjj strategies, we refer the reader to the methods proposed as future work in Section~\ref{sec:conclusions}.

\subsection{Investigation and Mitigation: U-Net Stacking}
\label{A:unet_stacking}

\begin{figure*}[h!]
    \centering
    \includegraphics[width=\textwidth]{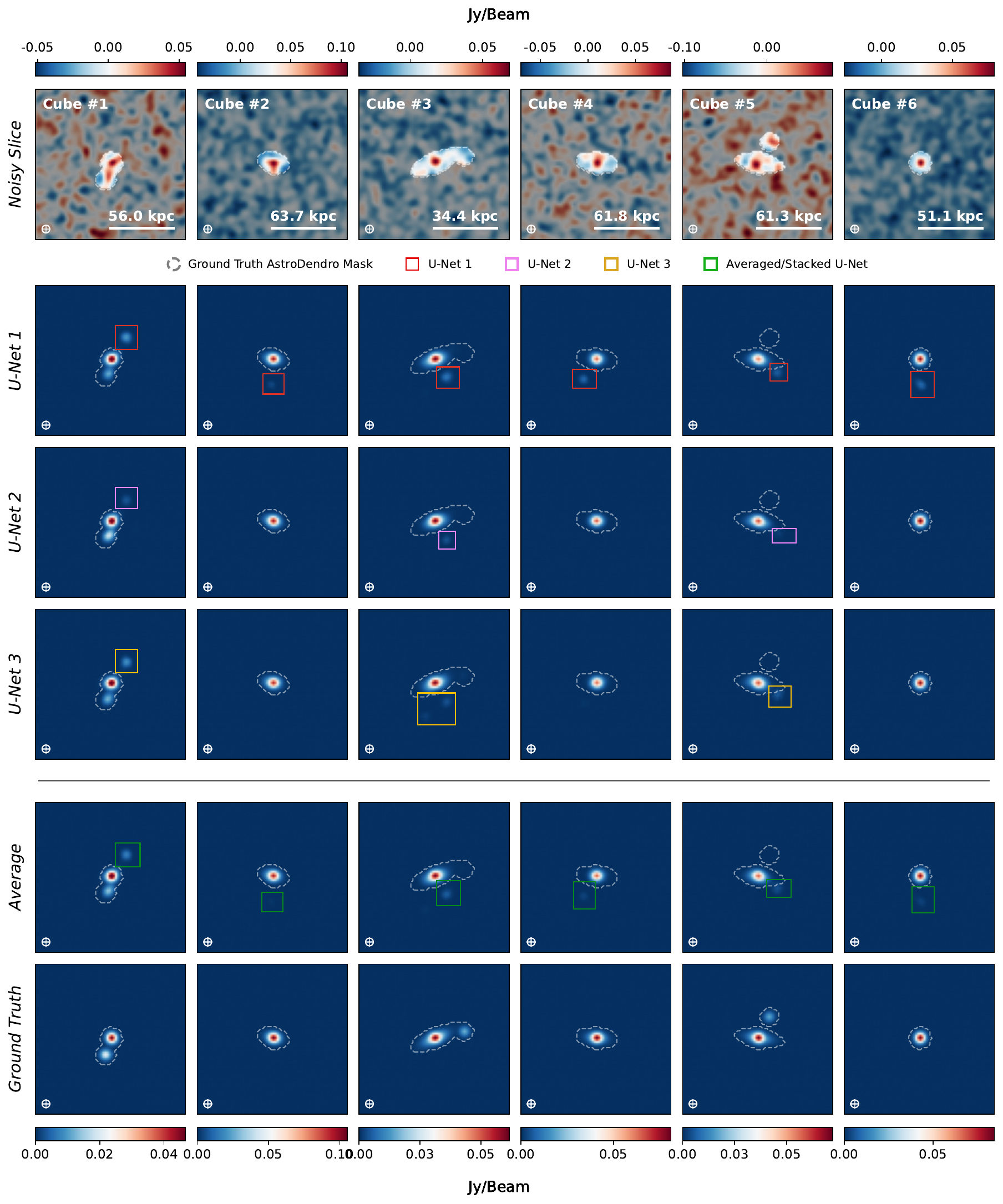}
    \caption{Visual analysis of hallucinated signal in the cubes analysed with our U-Net denoising model (U-Net 1) in Section \ref{sec:unet_hallucinations} and Figure \ref{fig:hallucinations}, extended here using two additional U-Nets (U-Nets 2 and 3). All three networks share the same initialisation parameters and were re-trained on the identical dataset. Hallucinations are marked and color-coded according to the U-Net that produced them.}
    \label{A:fig:unet_stacking}
\end{figure*}

To better understand the behaviour of our U-Net denoising models and the origins of hallucinated signals, we performed a detailed analysis using a \textbf{U-Net stacking approach}. This approach involves training multiple independent instances of the same network architecture - in this case, three U-Nets (U-Net 1, 2, and 3) - using identical initialisation parameters and the same training dataset. By comparing the outputs across these independently trained networks, we can probe the \textbf{stability, reproducibility, and overconfidence} of U-Net predictions, particularly in low signal-to-noise regions where hallucinations are more likely to occur.

U-Net stacking is useful for several reasons:
\begin{itemize}
    \item \textbf{Quantifying Model Overconfidence:} If hallucinations appear consistently in the same location across all models, this may indicate that the network has learned to assign high confidence to specific features, regardless of their presence in the true signal.
    \item \textbf{Assessing Non-Deterministic Behaviour:} Variations in hallucinations across models reveal the stochastic nature of network predictions due to random initialization, training dynamics, and optimization processes.
    \item \textbf{Mitigating Hallucinations:} Combining multiple models (ensembling) or using stacking can reduce spurious features, as inconsistent hallucinations tend to cancel out, highlighting only the robust, reproducible signal.
\end{itemize}

We applied this methodology to six cubes that exhibited hallucinations in U-Net 1 (the model used in the main analysis presented in Section \ref{sec:unet_hallucinations} and Figure \ref{fig:hallucinations}). The results can be summarized as follows:

\begin{itemize}
    \item \textbf{Cube 1 and 4:} Hallucinations appear at the same locations across all three U-Nets, although their amplitudes vary. This indicates a degree of \textbf{model overconfidence}, suggesting that the network consistently interprets certain low-SNR regions as signal even when they are not present.  
    
    \item \textbf{Cubes 2, 4, and 6:} Hallucinations present in U-Net 1 are completely mitigated in U-Nets 2 and 3. This demonstrates that these hallucinations are \textbf{non-deterministic} and may be attributable to the stochastic nature of training rather than inherent features in the dataset.  
    
    \item \textbf{Cube 3:} Additional or differing hallucinations appear in U-Nets 2 and 3 compared to U-Net 1, highlighting the \textbf{variability of hallucinations} across independently trained networks and further supporting the stochastic interpretation.
\end{itemize}

Figure \ref{A:fig:unet_stacking} provides a visual summary, where hallucinations are marked and color-coded according to the U-Net responsible for each spurious feature. This representation clearly illustrates both consistent overconfident predictions (e.g., Cube 1) and stochastic hallucinations that vary between networks (e.g., Cubes 2, 3, 4, and 6).

\end{document}